\documentclass{aa}
\usepackage{graphicx}
\usepackage{bm}
\usepackage{relsize}
\usepackage{amsmath}
\usepackage{mathtools}
\usepackage{threeparttable}
\newcommand{\snp}[1]{\hspace{-#1pt}}

\newcommand{\itover}[2]{\,\hspace{.3mm}#1{\!\hspace{-.3mm}#2}}
\newcommand{\itilde}[1]{\itover{\widetilde}{#1}}
\newcommand{\ithat}[1]{\itover{\hat}{#1}}

\newcommand{\itbar}[1]{\itover{\overline}{#1}}
\newcommand{\itbreve}[1]{\itover{\breve}{#1}}

\newcommand{\ovec}[1]{{\mbox{\boldmath $#1$}}}

\newcommand{\zervec}{\ovec{0}}
\newcommand{\ivm}{\phantom{-}}
\newcommand{\deriv}[3]{\frac{#3\hspace*{-.06em} {#1}}{#3\hspace*{.06em} {#2}}}
\newcommand{\derivil}[3]{{#3 {#1}/#3 {#2}}}

\newcommand{\parder}[2]{\deriv{#1}{#2}{\partial}}
\newcommand{\parderil}[2]{\derivil{#1}{#2}{\partial}}
\newcommand{\rezip}[1]{\frac{1}{#1}}

\newcommand{\myref}[1]{~\hspace{0pt plus 1pt minus 1pt}\ref{#1}}

\newcommand{\eqsref}[2]{Eqs.\ (\ref{#1}) and (\ref{#2})}

\newcommand{\sectref}[1]{Sect.\myref{#1}}

\newcommand{\figref}[1]{Fig.\myref{#1}}

\newcounter{saveqn}
\newcommand{\alpheqn}{\refstepcounter{equation}\setcounter{saveqn}{\value{equati
on}}%
\setcounter{equation}{0}%
\renewcommand{\theequation}{\mbox{\arabic{chapter}.\arabic{saveqn}\alph{equation
}}}}
 
\newcommand{\reseteqn}{\setcounter{equation}{\value{saveqn}}%
\renewcommand{\theequation}{\arabic{chapter}.\arabic{equation}}}

\newcommand{\EQ}{\begin{equation}}
\newcommand{\EN}{\end{equation}}
\newcommand{\EQA}{\begin{eqnarray}}
\newcommand{\ENA}{\end{eqnarray}}
\newcommand{\eq}[1]{(\ref{#1})}
\newcommand{\EEq}[1]{Equation~(\ref{#1})}
\newcommand{\Eq}[1]{Equation~(\ref{#1})}
\newcommand{\Eqs}[2]{Equations~(\ref{#1}) and~(\ref{#2})}

\newcommand{\App}[1]{Appendix~\ref{#1}}
\newcommand{\Sec}[1]{Section~\ref{#1}}

\newcommand{\Fig}[1]{Figure~\ref{#1}}
\newcommand{\FFig}[1]{Figure~\ref{#1}}
\newcommand{\Figs}[2]{Figures~\ref{#1} and \ref{#2}}

\newcommand{\Tab}[1]{Table~\ref{#1}}

{}
{}
\newcommand{\meanFFFF}{\overline{\mbox{\boldmath ${\cal F}$}}{}}{}

{}
{}
{}
\newcommand{\meanEEEE}{\overline{\mbox{\boldmath ${\cal E}$}}{}}{}
{}
{}
{}
{}
\newcommand{\meanAA}{\overline{\mbox{\boldmath $A$}}{}}{}
\newcommand{\meanBB}{\overline{\mbox{\boldmath $B$}}{}}{}
{}
{}
{}
{}
{}
{}
{}
\newcommand{\meanJJ}{\overline{\mbox{\boldmath $J$}}{}}{}
\newcommand{\meanUU}{\overline{\mbox{\boldmath $U$}}}

\newcommand{\meanWW}{\overline{\mbox{\boldmath $W$}}{}}{}
{}
{}

\newcommand{\meanB}{\overline{B}}

\newcommand{\meanP}{\overline{P}}

\newcommand{\meanJ}{\overline{J}}

\newcommand{\meanEEE}{\overline{\cal E}}
\newcommand{\meanFFF}{\overline{\cal F}}

\newcommand{\hatBB}{\hat{\bm{B}}}
{}

{}
{}

\newcommand{\hatB}{\hat{B}}

\newcommand{\bbN}{\bb_0} 
\newcommand{\uuN}{\uu_0} 
\newcommand{\wwN}{\ww_0} 
\newcommand{\jjN}{\jj_0}

\newcommand{\pN}{p_0}
\newcommand{\bbB}{\bb_{\hspace*{-1.1pt}\itbar{B}}}  
\newcommand{\jjB}{\jj_{\hspace*{-1.1pt}\itbar{B}}}
\newcommand{\uuB}{\uu_{\hspace*{-1.1pt}\itbar{B}}}
\newcommand{\wwB}{\ww_{\hspace*{-1.1pt}\itbar{B}}}

\newcommand{\pB}{p_{\hspace*{-1.1pt}\itbar{B}}}
\newcommand{\aaN}{\aaaa_0}
\newcommand{\aaB}{\aaaa_{\hspace*{-1.1pt}\itbar{B}}}
\newcommand{\EEEEN}{\meanEEEE_0}
\newcommand{\EEEEB}{\meanEEEE_{\hspace*{-1.1pt}\itbar{B}}}
\newcommand{\EEEN}{\EEE_0}
\newcommand{\EEEB}{\EEE_{\hspace*{-1.1pt}\itbar{B}}}

\newcommand{\FFFFBK}{\FFFFB^\sK}
\newcommand{\EEEEBK}{\EEEEB^\sK}
\newcommand{\FFFFBM}{\FFFFB^\sM}
\newcommand{\EEEEBM}{\EEEEB^\sM}
\newcommand{\FFFFBMK}{\FFFFB^\sMK}
\newcommand{\EEEEBMK}{\EEEEB^\sMK}

\newcommand{\FFFFN}{\meanFFFF_0}
\newcommand{\FFFFB}{\meanFFFF_{\hspace*{-1.1pt}\itbar{B}}}
\newcommand{\FFFN}{\FFF_0}
\newcommand{\FFFB}{\FFF_{\hspace*{-1.1pt}\itbar{B}}}

\newcommand{\smallT}{{\rm T}}

\newcommand{\meanEEEET}{\meanEEEE^{\smallT}}
\newcommand{\meanBBT}{\meanBB^{\smallT}}
\newcommand{\meanJJT}{\meanJJ^{\smallT}}
\newcommand{\bbT}{\bb^{\smallT}}
\newcommand{\aaT}{\aaaa^{\smallT}}
\newcommand{\uuT}{\uu^{\smallT}}
\newcommand{\wwT}{\ww^{\smallT}}
\newcommand{\jjT}{\jj^{\smallT}}
\newcommand{\pT}{p^{\smallT}}
\newcommand{\FFFT}{\FFF_{\smallT}}
\newcommand{\EEET}{\EEE_{\smallT}}
\newcommand{\FFFFT}{\meanFFFF^{\smallT}}
\newcommand{\EEEET}{\meanEEEE^{\smallT}}
\newcommand{\BBimp}{\BB_{\text{imp}}}
\newcommand{\Bimp}{B_{\text{imp}}}
\newcommand{\tensor}[1]{\boldsymbol{\mathsf #1}}
\newcommand{\aTens}{\tensor{\alpha}}
\newcommand{\bTens}{\tensor{\eta}}
\newcommand{\QK}{{\rm{QK}}}

\newcommand{\alphaQk}{\alpha^\QK}
\newcommand{\etatQk}{\etat^\QK}
\newcommand{\sM}{{\rm M}}
\newcommand{\sK}{{\rm K}}
\newcommand{\sMK}{{\rm R}}
\newcommand{\sm}{{\rm m}}
\newcommand{\sk}{{\rm k}}
\newcommand{\smk}{{\rm mk}}
\newcommand{\ei}[1]{{\rm e}^{{\rm i} #1}}
\newcommand{\phiTens}{\boldsymbol{\phi}}
\newcommand{\psiTens}{\boldsymbol{\psi}}
\newcommand{\zzz}{\hat{\mbox{\boldmath $z$}} {}}

\newcommand{\nullvector}{{\bf0}}

\newcommand{\ww}{\mbox{\boldmath $w$} {}}

\newcommand{\ppsi}{\mbox{\boldmath $\psi$} {}}
\newcommand{\ppsiN}{\ppsi_0}
\newcommand{\kk}{\bm{k}}

\newcommand{\uu}{\mbox{\boldmath $u$} {}}
\newcommand{\UU}{\mbox{\boldmath $U$} {}}

\newcommand{\bb}{\mbox{\boldmath $b$} {}}
\newcommand{\BB}{\mbox{\boldmath $B$} {}}

\newcommand{\jj}{\mbox{\boldmath $j$} {}}
\newcommand{\JJ}{\mbox{\boldmath $J$} {}}

\newcommand{\AAA}{\mbox{\boldmath $A$} {}}
\newcommand{\aaaa}{\mbox{\boldmath $a$} {}}

\newcommand{\ff}{\mbox{\boldmath $f$} {}}

\newcommand{\FF}{\mbox{\boldmath $F$} {}}
\newcommand{\EEE}{\mbox{\boldmath ${\cal E}$} {}}
\newcommand{\FFF}{\mbox{\boldmath ${\cal F}$} {}}

\newcommand{\WW}{\mbox{\boldmath $W$} {}}

\newcommand{\nab}{\mbox{\boldmath $\nabla$} {}}

\newcommand{\ii}{{\rm i}}

\newcommand{\curl}{{\rm curl} \, {}}
\newcommand{\dive}{{\rm div}  \, {}}

\def\Pm{\mbox{\rm Pr}_{\rm M}}
\def\Rm{\mbox{\rm Re}_{\rm M}}

\def\Rey{\mbox{\rm Re}}

\def\Lu{\mbox{\rm Lu}}

\def\yL{y_{\rm L}}
\def\yN{y_{\rm N}}

\def\kf{k_{\rm f}}

\def\rms{{\mathrm{rms}}}

\def\urms{u_{\rm rms}}
\def\urmsN{u_{0\rm rms}}

\def\brms{b_{\rm rms}}
\def\brmsN{b_{0\rm rms}}

\def\etat{\eta_{\rm t}}

\newcommand{\yr}{\,{\rm yr}}

\newcommand{\yapj}[3]{ #1, {ApJ,} {#2}, #3}

\newcommand{\yapjl}[3]{ #1, {ApJ,} {#2}, #3}

\newcommand{\yan}[3]{ #1, {Astron.\ Nachr.,} {#2}, #3}

\newcommand{\yana}[3]{ #1, {A\&A,} {#2}, #3}

\newcommand{\ygafd}[3]{ #1, {Geophys.\ Astrophys.\ Fluid Dyn.,} {#2}, #3}

\newcommand{\yjfm}[3]{ #1, {J.\ Fluid Mech.,} {#2}, #3}

\newcommand{\ypfb}[3]{ #1, {Phys.\ Fluids B,} {#2}, #3}

\newcommand{\yjetp}[3]{ #1, {Sov.\ Phys.\ JETP,} {#2}, #3}
\newcommand{\yphy}[3]{ #1, {Physica,} {#2}, #3}
\newcommand{\yaraa}[3]{ #1, {ARA\&A,} {#2}, #3}

\newcommand{\yrpp}[3]{ #1, {Rep. Prog.\ Phys.,} {#2}, #3}
\newcommand{\yprs}[3]{ #1, {Proc.\ Roy.\ Soc.\ Lond.,} {#2}, #3}

\newcommand{\ymn}[3]{ #1, {MNRAS,} {#2}, #3}

\newcommand{\ypre}[3]{ #1, {Phys.\ Rev.\ E,} {#2}, #3}

\newcommand{\yjour}[4]{ #1, {#2}, {#3}, #4}

\newcommand{\ybook}[3]{ #1, {#2} (#3)}

\graphicspath{{fig/}}

\begin{document}

\authorrunning{M. Rheinhardt and A. Brandenburg}
\titlerunning{Test-field method with MHD background}
\title{Test-field\snp{1} method\snp{1} for\snp{2} mean-field\snp{2} coefficients\snp{1} with\snp{2} MHD\snp{2} background}
\author{M. Rheinhardt\inst{1} \and A. Brandenburg \inst{1,2}
}

\institute{
NORDITA, AlbaNova University Center, Roslagstullsbacken 23,
SE-10691 Stockholm, Sweden
\and
Department of Astronomy, AlbaNova University Center,
Stockholm University, SE-10691 Stockholm, Sweden
}

\date{\today,~ $ $Revision: 1.154 $ $}

\abstract{}{%
The test-field method for computing turbulent transport coefficients from
simulations of hydromagnetic flows
is extended to the regime with a magnetohydrodynamic (MHD) background.
}{%
A generalized set of test equations is derived using both
the induction equation and a modified momentum equation.
By employing an additional set of auxiliary equations,
we derive linear equations
describing the response of the system to a set of prescribed test fields.
Purely magnetic and MHD backgrounds are emulated by applying
an electromotive force in the induction equation analogously
to the ponderomotive force in the momentum equation.
Both forces are chosen to have Roberts flow-like geometry.
}{%
Examples with an MHD background are studied where the previously
used quasi-kinematic test-field method breaks down.
In cases with homogeneous mean fields
it is shown that the generalized test-field method produces
the same results as the imposed-field method, where the field-aligned component
of the actual electromotive force from the simulation is used.
Furthermore, results for the turbulent diffusivity tensor are given,
which are inaccessible to the imposed-field method.
For MHD backgrounds, new mean-field effects are found that depend
on the occurrence of cross-correlations between magnetic and velocity
fluctuations.
For strong imposed fields, $\alpha$ is found to be quenched proportional
to the fourth power of the field strength, regardless of the type of
background studied.
}{}


\maketitle


\section{Introduction}

Astrophysical bodies such as stars with outer convective envelopes,
accretion discs, and galaxies tend to be magnetized.
In all those cases the magnetic field varies on a broad spectrum of scales.
On small scales the magnetic field might well be the result of scrambling an
existing large-scale field by a small-scale flow. 
However, at large magnetic Reynolds numbers, i.e.\ when advection dominates
over magnetic diffusion, another source of small-scale fields is
small-scale dynamo action (Kazantsev 1968).
This process is now fairly well understood and confirmed by numerous
simulations (Cho \& Vishniac 2000; Schekochihin et al.\ 2002, 2004;
Haugen et al.\ 2003, 2004); for a review see Brandenburg \& Subramanian (2005).
Especially in the context of magnetic fields of galaxies, the
occurrence of small-scale dynamos
may be important for providing a strong
field on short time scales ($10^7\yr$), which may then act as a seed
for the large-scale dynamo (Beck et al.\ 1994).

In contemporary galaxies the magnetic fields on small and large length scales
are comparable (Beck et al.\ 1996), but in stars this is less clear.
On the solar surface the solar magnetic field shows significant amounts
of small-scale fields (Solanki et al.\ 2006).
The possibility of generating such magnetic fields locally in the upper layers
of the convection zone by a small-scale dynamo is sometimes referred to as
surface dynamo (Cattaneo 1999; Emonet \& Cattaneo 2001;
V\"ogler \& Sch\"ussler 2007).
On the other hand, simulations of stratified convection with shear show that
small-scale dynamo action is a prevalent feature of the kinematic regime,
but becomes less important when the field is strong and has saturated
(Brandenburg 2005a; K\"apyl\"a et al.\ 2008).

An important question is then how the primary
presence of small-scale magnetic fields affects
the generation of large-scale fields
if these are the result of a dynamo process
that produces magnetic fields on scales large compared with those of the
energy-carrying eddies of the underlying and in general turbulent flow (Parker 1979)
via an instability.
A commonly used tool for studying these large-scale dynamos is mean-field
electrodynamics, where correlations of small-scale magnetic and velocity
fields are expressed in terms of the mean magnetic field and the mean velocity using
corresponding turbulent transport coefficients or their associated integral kernels 
(Moffatt 1978; Krause \& R\"adler 1980).
The determination of these coefficients
(e.g., $\alpha$ effect and turbulent diffusivity)
is the central task of mean-field dynamo theory.
This can be performed analytically, but usually only via approximations
which are hardly justified in realistic astrophysical situations
where the magnetic Reynolds numbers, $\Rm$, are large.

Obtaining turbulent transport coefficients from direct numerical simulations (DNS) offers a more
sustainable alternative as it avoids the restricting approximations
and uncertainties of analytic approaches.
Moreover, no assumptions concerning correlation properties of the turbulence need to be made,
because a direct ``measurement" of those properties is performed
in a physically consistent situation emulated by the DNS.
The simplest way to accomplish such a measurement is to include, in the DNS, an imposed large-scale (typically uniform)
magnetic field whose influence on the fluctuations of magnetic field and velocity
is utilized to infer some of the full set of transport coefficients.
We refer to this technique as the {\em imposed-field method}.

A more universal tool is offered by the test-field method (Schrinner et al.\ 2005, 2007),
which allows the determination of all wanted transport coefficients from a single DNS.
For this purpose the fluctuating velocity is taken from the DNS and inserted into a properly
tailored set of {\em test equations}.
Their solutions, the {\em test solutions} 
represent fluctuating magnetic fields as responses  to the interaction of the fluctuating velocity with a set of
properly chosen mean fields.
These mean fields will be called {\em test fields}.
For distinction from the test equations,
which are in general also solved by direct numerical simulation,
we will refer to the original DNS as the {\em main run}.
This method has been successfully applied to homogeneous turbulence with
helicity (Sur et al.\ 2008, Brandenburg et al.\ 2008a),  with shear and no
helicity (Brandenburg et al.\ 2008b), and with both (Mitra et al.\ 2009).

A crucial requirement on any test-field method is the independence
of the resulting transport coefficients on the strength
and geometry of the test fields.
This is immediately clear in the kinematic situation, i.e., if
there is no back-reaction of the mean magnetic field
on the flow.
Indeed, for given magnetic boundary conditions and a given value for
the magnetic diffusivity,
the transport coefficients must not reflect anything
else than correlation properties of the velocity field
which are completely determined by the hydrodynamics alone.
For this to be guaranteed the test equations
have to be linear and the test solutions have to be linear and
homogeneous in the test fields.

Beyond the kinematic situation the same requirement still holds,
although the flow is now modified by a mean magnetic field occurring in the main run.
(Whether it is maintained by external sources or generated by a dynamo process does
not matter in this context.)
Consequently, the transport coefficients are now functions
of this mean field.
It is no longer so obvious that under these
circumstances a test-field method with the aforementioned linearity and
homogeneity properties can be established at all.
Nevertheless, it turned out that the test-field method
developed for the kinematic situation gives consistent results
even in the nonlinear case without any modification
(Brandenburg et al.\ 2008c).
This method, which we will refer to as
``quasi-kinematic" is, however, restricted to situations in which the
magnetic fluctuations are solely a consequence of the mean magnetic field.
(That is, the primary or background turbulence is purely hydrodynamic.)

The power of the quasi-kinematic method was demonstrated based on
a simulation of an  $\alpha^2$ dynamo where the main run has reached saturation
with mean magnetic fields being Beltrami fields (Brandenburg et al.\ 2008c).
Magnetic and fluid Reynolds numbers up to 600 were taken into account,
so in some of the high $\Rm$ runs there was certainly small-scale dynamo action,
that is, a primary magnetic turbulence $\bbN$ should be expected.
Nevertheless, the quasi-kinematic method was found to
work reliably even for strongly saturated dynamo fields.
This was revealed by verifying
that the analytically solvable mean-field dynamo model
employing the values of $\alpha$ and turbulent diffusivity as derived from the saturated state of the main run
indeed yielded a vanishing growth rate. Very likely the small-scale dynamo had saturated on a low level
so the contribution to the mean electromotive force,
which was not taken into account by the quasi-kinematic method,
could not create a marked error.

Limitations of the quasi-kinematic test-field method were recently
pointed out by Courvoisier et al.\ (2010) and will be commented upon in
more detail in the discussion section.
Indeed, the purpose of our work is to propose a
generalized test-field method that allows for the presence of magnetic
fluctuations in the background turbulence.
Moreover, its validity range should cover dynamically 
effective mean fields, that is, situations in which the velocity and magnetic
field fluctuations are significantly affected by the mean field.

With a view to this generalization we
will first recall the mathematical justification of the quasi-kinematic
method and indicate the reason for its limited applicability (Sect.\ 2).
In Sect.\ 3 the foundation of the generalized method will be laid
down in the context of a relevant set of model equations.
In Sect.\ 4 results will be presented for various combinations of
hydrodynamic and magnetic backgrounds having Roberts-flow geometry.
The astrophysical relevance of our results and the connection with the
work of Courvoisier et al.\ (2010) will be discussed in Sect.\ 5.

\section{Justification of the quasi-kinematic test-field method and its limitation}
\label{quasikin}
In the following we split any relevant physical quantity $F$ into
mean and fluctuating parts, $\itbar{F}$ and $f$.
No specific averaging procedure will be adopted at this point;
we merely assume the Reynolds rules to be obeyed.
Furthermore, we split the fluctuations of magnetic field
and velocity, $\bb$ and $\uu$, into parts existing already
in the absence of a mean magnetic field, $\bbN$ and $\uuN$
(together they form the background turbulence), and parts
vanishing with $\meanBB$, denoted by $\bbB$ and $\uuB$.
We may split the mean
electromotive force $\meanEEEE= \overline{\uu\times\bb}$ likewise
and get
\begin{align}
\meanEEEE &= \EEEEN + \EEEEB\\[-2mm]
\intertext{with}
 \EEEEN &= \overline{\uuN\times\bbN},\\
 \EEEEB &= \overline{\uuN\times\bbB} +  \overline{\uuB\times\bbN}
+ \overline{\uuB\times\bbB}\,.  \label{EEEE}
\end{align}
Note that
we do not restrict $\bbB$ and $\uuB$, and hence also not $\EEEEB$
to a certain order in $\meanBB$.

In the present Section we assume that the background turbulence is
purely hydrodynamic, that is, $\bbN=\zervec$ and hence $\bb=\bbB$
or, in other words, the magnetic fluctuations $\bb$ are entirely a
consequence of the interaction of the velocity fluctuations  $\uu$
with the mean field $\meanBB$.

In a homogeneous medium, the induction equations for the total and mean
magnetic fields as well as for the magnetic fluctuations read
\begin{align}
\parder{\BB}{t} &=\eta\nabla^2\BB+\curl(\UU\times\BB), \\
\parder{\meanBB}{t} &=\eta\nabla^2\meanBB+\curl(\meanUU\times\meanBB +\meanEEEE), \\
\parder{\bb}{t} &=\eta\nabla^2\bb+\curl(
\meanUU\times\bb+\uu\times\meanBB +\EEE'), \label{dbbdt}
\end{align}
with $\EEE' = \uu\times\bb - \overline{ \uu\times\bb}$. 
The solution of the linear equation \eqref{dbbdt} for the fluctuations $\bb$, {\em considered
as a functional of $\uu$, $\meanUU$ and $\meanBB$}, is linear
and homogeneous in the latter and the same is true for
\begin{equation}
\EEEEB=\meanEEEE=\overline{\uu\times\bbB}=\overline{\uu\times\bb}\,.  \label{EEEE1}
\end{equation}

If the velocity is influenced by the mean field, that is, if $\uu$ and $\meanUU$
depend on $\meanBB$, $\meanEEEE$ considered
as a functional of $\meanBB$, $\meanEEEE\{\meanBB\}$, is of course nonlinear.
However, $\meanEEEE$, again considered  as a functional of $\uu$, $\meanUU$ and $\meanBB$, 
$\meanEEEE\{\uu,\meanUU,\meanBB\}$, is still linear in $\meanBB$.

The major task of mean-field theory consists now just in establishing
a linear and homogeneous functional relating $\meanEEEE$ and $\meanBB$.
Making the ansatz
\begin{equation} \meanEEEE=\aTens\meanBB - \bTens
\nab \meanBB, \label{Eansatz} 
\end{equation}
with $\nab \meanBB$ being the gradient tensor of the mean magnetic field,
this task coincides
with determining  the tensors $\aTens$ and $\bTens$, which are of course
functionals of $\uu$ and $ \meanUU$.  Because of linearity and
homogeneity we are entitled to employ for this purpose various {\em
arbitrary} vector fields  $\meanBBT$ (i.e., test fields) in place of
$\meanBB$ in \eqref{dbbdt}, keeping the velocity of course fixed.
Each specific assignment of $\meanBBT$ yields a corresponding $\bbT$
and via that an $\meanEEEET$ and it establishes (up to) three linear
equations for the wanted components of $\aTens$ and $\bTens$.  Hence, choosing
the number of test fields in accordance with the number of the tensor
components, and specifying the geometry of the test fields ``sufficiently
independent" from each other, the components of $\aTens$ and $\bTens$ can be
determined uniquely.  In doing so, the amplitude of the test fields
clearly drops out (Brandenburg et al.\ 2008b).

Is the result affected by the \emph{geometry} of the test fields?
An ansatz like \eqref{Eansatz} is in general not exhaustive, but
restricted in its validity to a certain class of mean fields, here
strictly speaking to stationary fields which change at most linearly
in space. Consequently, the geometry of the test fields is without
relevance just as long as they are taken from the class for which
the $\meanEEEE$ ansatz is valid, but not for other choices.

For many applications it will be useful to generalize the test-field method
such that all employed test-fields are harmonic functions of position, defined by one and the same
wavevector $\kk$. The turbulent transport coefficients can then be obtained as functions of $\kk$
and have to be identified with the Fourier transforms of integral kernels which define
the in general non-local relationship between
$\meanEEEE$ and $\meanBB$ (Brandenburg et al.\ 2008a).
Quite analogous, the in general also non-instantaneous  relationship between these quantities can be recovered by using harmonic functions of time for the test-fields.
The coeffcients, then depending on the angular frequency $\omega$, represent again Fourier transforms
of the corresponding integral kernels
(Hubbard \& Brandenburg 2009).

If $\uu$ and $\meanUU$ are taken from a series of main runs with a dynamically
effective mean field of, say,
fixed geometry, but from run to run differing
strength $\meanB$, $\aTens$ and $\bTens$ can be obtained as functions of
$\meanB$.  Thus, it is possible to determine the {\em quenched}
dynamo coefficients basically in the same way as in the kinematic
case, albeit at the cost of multiple numeric work.

Let us now relax the above assumption on the background turbulence
and admit additionally a primary {\em magnetic} turbulence $\bbN$.
For the sake of simplicity we will not deal here with $\EEEEN$,
so let us assume that it vanishes.
In the representation  \eqref{EEEE} of $\meanEEEE$
we now combine the first and last terms using $\uu=\uuN+\uuB$
and obtain 
\begin{equation}
\meanEEEE = \overline{\uu\times\bbB}
+ \overline{\uuB\times\bbN},
\label{EMFgeneral}
\end{equation}
differing from \eqref{EEEE1} by the additional contribution,
$\overline{\uuB\times\bbN}$.
The quasi-kinematic
method necessarily fails here even when modifying \eqref{dbbdt}
appropriately to form an equation for $\bbB$
as only the term $\overline{\uu\times\bbB}$ is provided.
Obviously, a valid
scheme must treat also $\uuB$ in a test-field manner similar to
$\bbB$.  The equation to be  employed for $\uuB$ has of course to
rely upon the momentum equation.  Due to its intrinsic nonlinearity,
however,
a major challenge consists then in ensuring the linearity and
homogeneity of the test solutions in the test fields. 

\section{A model problem}

\subsection{Motivation}

We commence our study with a model problem
that is simpler than the complete MHD setup, but nevertheless shares with it the same mathematical
complications.
We drop the advection and pressure terms and adopt 
for the diffusion operator simply the Laplacian (and a homogeneous medium).
Thus, there is no constraint on the velocity from a continuity equation and an equation of state.
However, as in the full problem, we allow the magnetic field to exert a Lorentz
force on the fluid velocity.
We also allow for the presence of
an imposed uniform magnetic field $\BBimp$ to enable a
determination of the $\alpha$ effect independently from the test-field method.
The magnetic field is hence represented as $\BB=\BBimp+\nab\times\AAA$ where $\AAA$ is the  vector potential of its non-uniform part.
The resulting modified momentum equation for the velocity $\UU$ and the (original) induction equation read then
\begin{align}
\parder{\UU}{t} &=\JJ\times\BB+\FF_{\rm K}+\nu\nabla^2\UU, 
\label{dUdt}\\
\parder{\AAA}{t}&=\UU\times\BB+\FF_{\rm M}+\eta\nabla^2\AAA,
\label{dAdt}
\end{align}
where we have included the possibility of both kinetic and magnetic forcing
terms, $\FF_{\rm K}$ and $\FF_{\rm M}$,  respectively.
(In this paper we use hydrodynamic and kinetic forcing synonymously.)
We have adopted a system of units in which $\BB$ has the dimension of velocity
namely $\BB \coloneqq \BB/\sqrt{\mu\rho}$, $\rho=$const.
Defining the current density as $\JJ=\nab\times\BB$,
it has the unit of inverse time.
The electric field has then the unit of squared velocity.
Furthermore, $\nu$ is the kinematic viscosity and $\eta$ the magnetic
diffusivity.

As will become clear, the major difficulty in defining a test-field method for MHD or purely magnetic background turbulence
is caused by bilinear (or quadratic) terms like $\JJ\times\BB$ and $\UU\times\BB$. Hence, taking the $\UU\cdot\nab\UU$ nonlinearity into account
would not offer any new aspect, but would blur the essence of the derivation and the clear analogy
in the treatment of the former two nonlinearities.
The treatment of the advective term follows the same pattern, as is
demonstrated in \App{IsothermalProblem}.
Also, it should be pointed out that
working with the simpler set of equations helps reducing the
risk of errors in the numerical implementation.

In three dimensions and for $\BBimp=\FF_{\rm M}=\nullvector$, but  helical or non-helical kinetic forcing 
via $\FF_{\rm K}$, this system of equations is capable of reproducing
essential features of turbulent dynamos like initial exponential growth
and subsequent saturation; see, e.g., Brandenburg (2001) or
Haugen et al.\ (2004).

If $\BBimp\neq\nullvector$ or $\FF_{\rm M}\neq\nullvector$ we are
no longer dealing with a dynamo problem in the strictest sense.
A discussion of dynamo processes can still be meaningful if $\BBimp=\nullvector$ and the magnetic forcing
does not give rise to a mean electromotive force $\EEEN$. 
A possibility to accomplish this is $\FF_{\rm K}=\nullvector$
together with a magnetic forcing resulting in a Beltrami field $\bbN$, but any choice providing an isotropic background turbulence $\bbN$, $\uuN$
should be suited likewise.

Note that the mean-field induction equation is still autonomous
allowing for the solution $\meanBB=\nullvector$. It then depends on properties of the background turbulence like
chirality whether, e.g., the $\alpha$ effect or other mean-field effects render this solution unstable by enabling
growing solutions.

If we admit $\BBimp\neq\nullvector$, the two forcing terms can in
principle be adjusted such that the resulting background turbulence is
again isotropic and $\EEEEN=\nullvector$.
Should a growing mean field then be observed, it can legitimately
be attributed to an instability as $\meanBB=\BBimp$ is a solution
of the mean field induction equation and the imposed field cannot grow.
Thus both scenarios have the potential to exhibit  \emph{mean-field} dynamos although the original induction equation is inhomogeneous
and the dynamo must not  be considered as an instability of the completely non-magnetic state.

Models of the latter type may well have astrophysical relevance, because
at high magnetic Reynolds numbers small-scale dynamo action is expected
to be ubiquitous.
Large-scale fields are still considered to be the consequence of an instability,
at least if there is no $\EEEEN$ or any other sort of ``battery effect".
Magnetic forcing can be regarded as a modeling tool for providing a magnetic
background turbulence when, e.g., in a DNS the conditions
for small-scale dynamo action are not afforded.

More generally, magnetic forcing and  an imposed field provide excellent means of studying the $\alpha$
effect, the inverse cascade of magnetic helicity, and flow properties
in the magnetically dominated regime (see, e.g., Pouquet et al.\ 1976;
Brandenburg et al.\ 2002; Brandenburg \& K\"apyl\"a 2007).

\subsection{Purely magnetic background turbulence}
Before taking on the most general situation of both magnetic and velocity
fluctuations in the background turbulence it seems instructive to look
first at the case complementary to that discussed in \Sec{quasikin}.
That is, we
assume, perhaps somewhat artificially, that the background velocity 
fluctuations vanish, i.e.\ $\uuN=\zervec$, so that $\uu=\uuB$.
According to \eqref{EEEE} we now find
\begin{equation}
\meanEEEE = \EEEEB = \overline{\uuB\times\bb} = \overline{\uu\times\bb}.
\label{PurelyMagnetic}
\end{equation}
The modified momentum equation for the velocity fluctuations in a
homogeneous medium reads (cf. \eqref{dUdt})
\begin{equation}
\parder{\uu}{t}
=\meanJJ\times\bb+\jj\times\meanBB+\FFF'+\nu\nabla^2\uu,
\label{duudt}
\end{equation}
with $\FFF' = \jjB\times\bb+\jjN\times\bbB - \overline{\jjB\times\bb}+ \overline{\jjN\times\bbB}$. 
As  $(\jjN\times\bbN)'$ needs to vanish in order to guarantee $\uuN=\zervec$, this could
also be written as $\FFF' = \jj\times\bb - \overline{ \jj\times\bb}$.
Unlike in the quasi-kinematic method there is now no longer any way to base a test-field method
upon considering one of the fluctuating fields, here $\bb$, to be given (e.g.\ taken from a main run) while interpreting
the other, here $\uu$, as a linear and homogeneous functional of the mean field.
(This would work here, however, in the second order correlation approximation, where $\FFF'$ is set to zero.)

\subsection{General mean-field treatment}
\label{GeneralTreatment}

The mean-field equations for $\meanUU$ and $\meanBB=\curl\meanAA+\BBimp$ obtained by averaging
\eqsref{dUdt}{dAdt} are
\begin{align}
{\partial\meanUU\over\partial t}&=\nu\nabla^2\meanUU+\meanJJ\times\meanBB
+\meanFFFF,
\label{dUmdt}\\
{\partial\meanAA\over\partial t}&=\eta\nabla^2\meanAA+\meanUU\times\meanBB
+\meanEEEE,
\label{dAmdt}
\end{align}
where we have assumed that the mean forcing terms vanish.
From now on we extend our considerations also onto the relation between the
mean force $\meanFFFF=\overline{\jj\times\bb}$ and
the mean field.
In full analogy to the mean electromotive force we find this relation, considered
as a functional of $\uu$, $\meanUU$ and $\meanBB$, again to be linear and homogeneous in the latter
and write, to start with,
\begin{equation}
\FFFFB=\boldsymbol{\phi} \meanBB - \boldsymbol{\psi} \nab \meanBB.\label{Fansatz}
\end{equation}
Like $\aTens$ and $\bTens$,
the tensors $\boldsymbol{\phi}$ and $\boldsymbol{\psi}$ may depend on $\meanB$.
In the kinematic limit $\boldsymbol{\phi}$ and $\boldsymbol{\psi}$
are expected to be non-vanishing only if $\bbN\ne\zervec$.
An analysis in SOCA, however, would also require $\uuN\ne\zervec$ to get
a non-vanishing result; see \App{phideriv}.
Note that $\bbN\ne\zervec$ allows $\FFFFB$ to be linear in $\meanBB$,
which would otherwise be quadratic to leading order.
Consequently, the backreaction of the mean field onto the flow is no longer
independent of its sign.

As $\FFFFB$ is the divergence of the mean Maxwell tensor,
it has to vanish in the homogeneous case, i.e.\ for
homogeneous turbulence and a uniform field.
Hence, for \Eq{Fansatz} to be valid in physical space, $\boldsymbol{\phi}$
has then to vanish.
However, in Fourier space we may retain relation \eqref{Fansatz} with 
$\lim_{\kk\rightarrow\zervec} \boldsymbol{\phi}(\kk) = \zervec$
(but not so for $\psiTens$).
On the other hand, in the physical space 
a description of $\FFFFB$ employing the second derivative of $\meanBB$
is likely to be more appropriate, i.e.\
\begin{equation}
\FFFFB=\phiTens^*\nab^2\meanBB-\psiTens\nab\meanBB.
\label{Fansatz2}
\end{equation}
According to the expression for $\boldsymbol{\phi}(\kk)$,
which is derived in \App{phideriv} for Roberts forcing,
\Eq{Fansatz2}
specified to 
\[
\FFFFB=\phiTens^*\parder{\,^2\meanBB}{z^2}-\psiTens\meanJJ
\]
would indeed be sufficient as long as there is sufficient
scale separation between mean and fluctuating fields.
In the following, we continue referring to $\phiTens$ as introduced by
\Eq{Fansatz}.

The equations for the fluctuations are obtained by subtracting
\eq{dUmdt} from \eq{dUdt}, and \eq{dAmdt} from \eq{dAdt}, what leads
to
\begin{equation}
{\partial\uu\over\partial t}
=\meanJJ\times\bb+\jj\times\meanBB
+\FFF'+\ff_{\rm K}+\nu\nabla^2\uu,
\label{dudt}
\end{equation}
\begin{equation}
{\partial\aaaa\over\partial t}
=\meanUU\times\bb+\uu\times\meanBB
+\EEE'+\ff_{\rm M}+\eta\nabla^2\aaaa,
\label{dadt}
\end{equation}
respectively, where $\FFF'=\jj\times\bb - \overline{\jj\times\bb}$ and
$\EEE'=\uu\times\bb-\overline{\uu\times\bb}$
are terms that are quadratic in the
correlations, while $\ff_{\rm K,M}$
are just the fluctuating parts of the forcing functions.

To arrive at a set of equations that are formally linear and allow for solutions
as responses to a given mean field that are formally linear and homogeneous in the latter
we make use of the split of all quantities into parts existing in the absence 
of $\meanBB$ and parts vanishing with  $\meanBB$ introduced in \sectref{quasikin}.
We write $\uu=\uuN+\uuB$, $\aaaa=\aaN+\aaB$ and $\jj=\jjN+\jjB$, further
$\FFF'=\FFFN'+\FFFB'$ and $\EEE'=\EEEN'+\EEEB'$
and assume that the forcing is independent of $\meanBB$.
\EEq{dudt} and \eqref{dadt} split consequently as follows
(see \App{Linear} for an illustration)
\begin{align}
{\partial\uuN\over\partial t}
&=\nu\nabla^2\uuN+\FFFN'+\ff_{\rm K},
\label{du0dt}\\
{\partial\aaN\over\partial t}
&=\eta\nabla^2\aaN+\meanUU\times\bbN+\EEEN'+\ff_{\rm M},
\label{da0dt}
\end{align}
\begin{align}
{\partial\uuB\over\partial t}
&=\nu\nabla^2\uuB+\meanJJ\times\bb+\jj\times\meanBB
+\FFFB',
\label{du1dt}\\
{\partial\aaB\over\partial t}
&=\eta\nabla^2\aaB+\meanUU\times\bbB+\uu\times\meanBB
+\EEEB'.
\label{da1dt}
\end{align}
Because of $\FFFN'=(\jjN\times\bbN)'$ and $\EEEN'=(\uuN\times\bbN)'$,
\eqsref{du0dt}{da0dt} are completely closed. Furthermore, we have
\begin{align}
\FFFB'&=(\jjN\times\bbB+\jjB\times\bbN+\jjB\times\bbB)',
\label{FFF1}\\
\EEEB'&=(\uuN\times\bbB+\uuB\times\bbN+\uuB\times\bbB)'.
\label{EEE1}
\end{align}
We can rewrite these expressions such that they become formally linear in
$\uuB$ and $\bbB$ each in two different flavors:
\begin{alignat}{2}
\FFFB'&=(\jj\times\bbB+\jjB\times\bbN)'&&=(\jjN\times\bbB+\jjB\times\bb)',\label{FFFBd}\\
\EEEB'&=(\uu\times\bbB+\uuB\times\bbN)'&&=(\uuN\times\bbB+\uuB\times\bb)'.\label{EEEBd}
\end{alignat}
Now we have achieved our goal of deriving a system of formally linear equations
defining  the part of the
fluctuations that can be related to the mean field as response to the interaction 
with the given fluctuating fields $\uu$, $\uuN$, $\bb$, and $\bbN$.

Splitting mean force and electromotive force we find
\begin{equation}
\FFFFN=\overline{\jjN\times\bbN}\quad\mbox{and}\quad
\EEEEN=\overline{\uuN\times\bbN}
\end{equation}
for the parts existing already with $\meanBB=\nullvector$, due to a small-scale
dynamo or magnetic forcing.
Although it is hard to imagine that isotropic forcing alone is capable of enabling a
non-vanishing $\EEEEN$, an additional vector influencing the otherwise isotropic
turbulence may well act in this way.
For example,
using the second-order correlation approximation (SOCA) it was found that
in the presence of a non-uniform
mean flow $\meanUU$, with mean vorticity $\meanWW=\curl\meanUU$, we have
in  ideal MHD ($\eta=\nu=0$)
\begin{equation}
\EEEEN = -\frac{\tau_U}{3}\,\overline{\uuN\cdot\jjN} \,\meanUU +
                  \frac{\tau_W}{3}\,\overline{\uuN\cdot\bbN} \, \meanWW. \label{E0ansatz}
\end{equation}
Here the index "0" refers to the fluctuating background uninfluenced by both the magnetic field and the mean flow.
Beyond this specific result, too, one may expect that some cross correlation
of the primary turbulences is crucial.
(Yoshizawa 1990; R\"adler \& Brandenburg 2010).

For the parts vanishing with $\meanBB$ we have
\begin{alignat}{2}
\FFFFB&=\overline{\jj\times\bbB}+\overline{\jjB\times\bbN} &&=\overline{\jjN\times\bbB}+\overline{\jjB\times\bb},\label{FFFB}\\
\EEEEB&=\overline{\uu\times\bbB}+\overline{\uuB\times\bbN} &&=\overline{\uuN\times\bbB}+\overline{\uuB\times\bb}.\label{EEEB}
\end{alignat}
We recall that for $\bbN=\zervec$ (see \Sec{quasikin}), 
only the term $\overline{\uu\times\bbB}\equiv\EEEEBK$ occurs in the mean electromotive force
and for $\uuN=\zervec$ (see  \Eq{PurelyMagnetic}) only $\overline{\uuB\times\bb}\equiv\EEEEBM$.
For interpretation purposes, it is therefore convenient to 
define a correspondingly symmetrized version,
\begin{alignat*}{2}
\FFFFB&=\overline{\jj\times\bbB}+\overline{\jjB\times\bb}
-\overline{\jjB\times\bbB}&&=\FFFFBK+\FFFFBM+\FFFFBMK\\[4pt]
\EEEEB&=\overline{\uu\times\bbB}+\overline{\uuB\times\bb}
-\overline{\uuB\times\bbB}&&=\EEEEBK+\EEEEBM+\EEEEBMK,   
\end{alignat*}
with $\FFFFBMK=-\overline{\jjB\times\bbB}$ and
$\EEEEBMK=-\overline{\uuB\times\bbB}$ being
residual terms.  Of course this split is only meaningful with a non-vanishing mean field in the main run.
The corresponding transport coefficients might be split analogously.
Note, however, that for an imposed field in, say, the $x$ direction this is restricted to the $(1j)$ components of the tensors.
 
\subsection{Test-field method}
In a next step we define the actual test equations starting from Eqs. \eqref{du1dt}, \eqref{da1dt}, \eqref{FFFBd} and \eqref{EEEBd}.
As they are already arranged to be formally linear when  deliberately ignoring the relations between $\uuB$ and $\uu$ as well as between $\bbB$ and $\bb$, respectively,
we have nothing more to do than to identify $\meanBB$ with a test field $\meanBBT$ and $\bbB$,$\uuB$ with the corresponding test solutions $\bbT$,$\uuT$.
Due to the ambiguity in \eqref{FFFBd} and \eqref{EEEBd} four different versions are obtained reading
\begin{align}
{\partial\uuT\over\partial t}
&=\meanJJT\times\bb+\jj\times\meanBBT
+\FFF_T'+\nu\nabla^2\uuT,
\label{duTdt}\\
{\partial\aaT\over\partial t}
&=\meanUU\times\bbT+\uu\times\meanBBT
+\EEE_T'+\eta\nabla^2\aaT,
\label{daTdt}
\end{align}
with
\begin{align}
\FFFT'&=\left\{\begin{aligned}
                    (&\jj\times\bbT+\jjT\times\bbN)'\\
                    &\mbox{or}\\
                    (&\jjN\times\bbT+\jjT\times\bb)',
                \end{aligned}\right.
\label{FTd}\\[2mm]
\EEET'&= \left\{\begin{aligned}
                   (&\uu\times\bbT+\uuT\times\bbN)'\\
                    &\mbox{or}\\
                   (&\uuN\times\bbT+\uuT\times\bb)'.
                   \end{aligned}\right.
\label{ETd}
\end{align}
For mean force and electromotive force expressed by the test solutions we write correspondingly
\begin{align}
\FFFFT&=\left\{\begin{aligned}
                    &\overline{\jj\times\bbT}+\overline{\jjT\times\bbN}\\
                    &\mbox{or}\\
                    &\overline{\jjN\times\bbT}+\overline{\jjT\times\bb},
                \end{aligned}\right.
\label{FT}\\[2mm]
\EEEET&= \left\{\begin{aligned}
                   &\overline{\uu\times\bbT}+\overline{\uuT\times\bbN}\\
                   &\mbox{or}\\
                   &\overline{\uuN\times\bbT}+\overline{\uuT\times\bb},
                   \end{aligned}\right.
\label{ET}
\end{align}
and stipulate that the choice within \eqsref{FT}{ET} is always to correspond to the choice in \eqsref{FTd}{ETd}.
As we will make use of all four possible versions we label them
in a unique way by the names of the fluctuating fields of the main run
that enter the expressions for $\FFF_T'$ and $\EEE_T'$.
Accordingly, we find by inspection of \eqsref{FTd}{ETd}
for the labels the combinations {\sf ju}, {\sf jb}, {\sf bu} and {\sf bb};
see \Tab{TableVersions}.

\begin{table}[t]\caption{
Illustration of generating the four versions of the
generalized test-field method by combining
the different ways of representing $\FFF_T'$ and $\EEE_T'$ in \eqsref{FTd}{ETd}}.
\vspace{2pt}\centerline{\begin{tabular}{c|c|c}
& $\jj\times\bbT+\jjT\times\bbN$ & $\jjN\times\bbT+\jjT\times\bb$ \\[3pt]
\hline\\[-6pt]
$\uu\times\bbT+\uuT\times\bbN$ & {\sf ju} & {\sf bu} \\
$\uuN\times\bbT+\uuT\times\bb$ & {\sf jb} & {\sf bb} 
\label{TableVersions}\end{tabular}}\end{table}

Now we conclude that for given $\uu$, $\bb$, $\uuN$, $\bbN$ and $\meanUU$ the test solutions  $\uuT$, $\bbT$ are linear and homogeneous
in the test fields $\meanBBT$ and that the same holds for $\FFFFT$ and $\EEEET$. Hence, the tensors $\aTens$,  $\bTens$, $\boldsymbol{\phi}$ and
$\boldsymbol{\psi}$ derived from these quantities will not depend on the test fields, but exclusively reflect properties of the given fluctuating fields and the mean velocity.
If these are affected by a mean field in the main run the tensor components will show a dependence on $\meanB$.
Thus, like in the quasi-kinematic method, quenching behavior can be identified.
We observe further that when using the mean field from the main run as
one of the test fields, the corresponding test solutions $\bbT$ and $\uuT$
will coincide with $\bb-\bbN$ and $\uu-\uuN$, respectively.

Summing up, we may claim that the presented generalized test-field method in either shape satisfies certain necessary conditions for the correctness of  its results.
But can we be confident, that these are sufficient, too? An obvious complication lies in the fact, that in contrast to the quasi-kinematic method
yielding the transport coefficients uniquely, we have now to deal with four different versions which need not be completely equivalent. 
Indeed we will demonstrate that the reformulation of the original problem into \eqsref{duTdt}{daTdt} with \eqsref{FTd}{ETd} introduces occasional spurious instabilities. 
As we presently see no strict mathematical argument for the identity of the outcomes of all four versions, we resort to an empirical justification of our approach.
This is what the rest of this paper is devoted to.
\paragraph{Remarks:}
(i) Applying the second order correlation approximation (SOCA) to the system
\eqref{duTdt}, \eqref{daTdt}, that is, neglecting $\FFF_T'$ and $\EEE_T'$,
melts the four versions down to one and thus removes all complications.\\
(ii) In the limit $\meanBB\rightarrow\zervec$ we have simultaneously $\uu\rightarrow\uuN$ and $\bb\rightarrow\bbN$, so again only one version remains.
The method has then of course no longer any value for quenching
considerations, but it still might be useful to overcome the limitations
of SOCA. \\
(iii) For $\bbN=\zervec$ the $\aaT$ equation \eqref{daTdt} with the first version of \eqref{ETd}, i.e.
\begin{equation}
\EEE_T'=(\uu\times\bb_T)' 
\label{FTandETstandard}
\end{equation}
and correspondingly
$\meanEEEE_T=\overline{\uu\times\bb_T}$,
but \eqref{duTdt} ignored, reverts to the quasi-kinematic method.
For comparison we will occasionally apply this method even when $\bbN\ne\zervec$
and label the quantities calculated in this way with an upper index ``QK".

\vspace{3mm}
From now on we define mean fields by averaging over
two directions, here over the $x$ and $y$ directions, that is, all
mean quantities depend merely on $z$ (if at all) and we get a 1D mean-field dynamo problem.
As a consequence,
$\meanB_z$ is constant  and there are only two non-vanishing components of 
$\nab \meanBB$, namely $\meanJ_{x}$ and  $\meanJ_{y}$
so only the evolution of $\meanB_{x}$ and $\meanB_{y}$ has to be considered.
Moreover, $\meanEEE_z$ is without influence on the evolution of $\meanBB$.
Hence, instead of \eqsref{Fansatz}{Eansatz} we can write
\begin{equation}
\meanFFF_i=\phi_{ij}\meanB_j-\psi_{ij}\meanJ_j,\quad
\meanEEE_i=\alpha_{ij}\meanB_j-\eta_{ij}\meanJ_j,
\label{FFFiandEEEi}
\end{equation}
where the original rank-three tensors $\boldsymbol{\psi}$ and $\bTens$ are degenerated to rank-two ones.

Only the four components of either tensor with $i,j=1,2$ are of interest, thus 
altogether 16 components need to be determined.
As one test field $\meanBBT$ comprises two relevant components and yields  an $\FFFFT$ and an $\EEEET$, each again with two relevant components,
 we need to consider solutions of \eqsref{du1dt}{da1dt}  with \eqsref{FTd}{ETd}
for a set of four different test fields.

\paragraph{Selection of test fields:}
The simplest choice are homogeneous fields in the $x$ and $y$ directions,
but they are only suited to determine the $\boldsymbol{\alpha}$ tensor.

All four tensors can be extracted by use of
test fields with either the $x$
or the $y$ component proportional to either $\cos kz$ or to $\sin kz$
and the other vanishing
(see, e.g., Brandenburg 2005b; Brandenburg et al.\ 2008a,
2008b; Sur et al.\ 2008).
That is, $\meanBBT$ is either $B^{p\rm c}_i=\delta_{ip}\cos kz$ or
$B^{p\rm s}_i=\delta_{ip}\sin kz,\, p=1,2$, where the superscript
$pq,\, q ={\rm c,s}$ labels the test field and the subscript
$i$ refers to its components.
By varying the wavenumber $k$, the wanted tensor components can in principle
be determined as functions of $k$, but are then no longer allowed to be
interpreted in the usual way (cf.\ Brandenburg et al.\ 2008a).
Here we refrain from doing so and fix $k$ to the smallest possible value $k=2\pi/L_z$
where $L_z$ is the extent of the computational domain in $z$ direction.
But even then we introduce some errors 
because the harmonic test fields do not belong to the class of mean fields for which the ansatzes  \eqsref{Fansatz}{Eansatz}
are exhaustive (see \sectref{quasikin}). We must be aware that the tensors calculated in this way are ``polluted" by some contributions
from terms with higher derivatives of $\meanBB$ in $\meanEEEE$ and $\meanFFFF$.
To monitor these departures we compare
the $\aTens$ and $\boldsymbol{\phi}$ tensors found with harmonic test fields with those obtained with uniform ones.

For each pair of test fields $(\BB^{p\rm c},\BB^{p\rm s})$ we determine $2\times4$ unknowns
by solving the linear systems
\begin{equation}
\meanFFF^{pq}_i=\phi_{ij}\meanB^{pq}_j-\psi_{ij}\meanJ^{pq}_j,\quad
\meanEEE^{pq}_i=\alpha_{ij}\meanB^{pq}_j-\eta_{ij}\meanJ^{pq}_j.
\end{equation}
$q=\,$c,s. Note that there is no coupling between the systems for $p=1$ and $p=2$.
Inversion of the rotation matrix
\EQ
\tensor{R}=\begin{pmatrix}
\cos kz & -\sin kz\\
\sin kz & \phantom{-}\cos kz
\end{pmatrix}
\EN
(with the angle $kz$) provides the solutions explicitly, hence we have
\EQ
\begin{pmatrix}
\phi_{ij}\phantom{k}\\
\psi_{ij}k
\end{pmatrix} 
=
\tensor{R}^{\mbox{t}}
\begin{pmatrix}
\meanFFF_i^{jc}\\
\meanFFF_i^{js}
\end{pmatrix},
\quad
\begin{pmatrix}
\alpha_{ij}\phantom{k}\\
\eta_{ij}k
\end{pmatrix}
=
\tensor{R}^{\mbox{t}}
\begin{pmatrix}
\meanEEE_i^{jc}\\
\meanEEE_i^{js}
\end{pmatrix}.
\label{mean_est_E_standard}
\EN
Here the superscript ``t" indicates transposition.

\subsection{Forcing functions, computational domains and boundary conditions}
\label{forcing}

For testing purposes, a common and convenient choice is the
Roberts flow forcing function,
\begin{equation}
\ff=\sigma\kf\Psi\zzz+\nab\times(\Psi\zzz)
\quad\mbox{with}\quad \Psi=\cos k_xx\cos k_yy,\label{robForce}
\end{equation}
and the effective forcing wavenumber $\kf=(k_x^2+k_y^2)^{1/2}$.
With the chosen averaging the Roberts forcing is isotropic in the $xy$ plane.
Furthermore, $\sigma$ is a parameter controlling the helicity of the flow:
with $\sigma=0$ it is non-helical
while for $\sigma=1$ it reaches maximum helicity.
If not declared otherwise, we will employ just maximally helical Roberts forcing.

The Roberts forcing function
can be employed for kinetic as well as magnetic
forcing, so we write $\ff_{\rm K,M}=N_{\rm K,M}\ff$,
where the $N_{\rm K,M}$ are amplitudes having
the units of acceleration and velocity squared, respectively.
Note that for $\sigma=1$ Eq.\ \eqref{robForce} yields a Beltrami field, i.e., it
has the property $\curl\ff=\kf\ff$.
Therefore, provided $\BBimp=\zervec$, kinetic and magnetic forcing act
completely uninfluenced from each other
and create just a flow and a magnetic field having exactly the Roberts geometry.
This is not the case for $\sigma\neq1$. 

The computational domain is a cuboid with quadratic base $L_x=L_y=2\pi$
while its $z$ extent remains adjustable and depends on the
smallest wavenumber in the $z$ direction, $k_{1z}$, to be considered.
However, as the Roberts forcing function is not $z$ dependent,
the runs from which only $\aTens$
and $\boldsymbol{\phi}$ are extracted were carried out in 2D with $k_{1z}=0$.

We choose here $k_x=k_y=k_1$
where $k_1=1$ is the smallest wavenumber that fits into $x$ and $y$ extent of the computational domain.
For random forcing the domain is always cubic, i.e.\ $L_x=L_y=L_z=2\pi$.
In all cases we assume periodic boundary conditions in all directions.
The results presented below are based on revision \texttt{r13439} of the
{\sc Pencil Code}\footnote{\texttt{http://pencil-code.googlecode.com}},
which is a modular high-order code (sixth order in space and third-order
in time) for solving a large range of different partial differential
equations.

\subsection{Control parameters and non-dimensionalization}

In cases with an imposed magnetic field, we set $\BBimp=(B_0,0,0)$.
Along with it the forcing amplitudes $N_{\rm K,M}$ are the most relevant control parameters.
The only remaining one is the magnetic Prandtl number, $\Pm=\nu/\eta$.

It is convenient to measure length in units of the inverse minimal wavenumber $k_1$,
time in units of $1/\eta k_1^2$, velocity in units of $\eta k_1$,
and the magnetic field also in units of $\eta k_1$.
The forcing amplitudes $N_{\rm K,M}$ are given in units of $\eta^2k_1^3$
and $\eta^2k_1^2$, respectively.
Results will also be presented in dimensionless form: $\alpha_{ij}$
and $\psi_{ij}$ in units of $\eta k_1$, $\eta_{ij}$ in units of
$\eta$, $\phi_{ij}$ in units of $\eta k_1^2$ if not declared otherwise.
Dimensionless quantities are denoted by a tilde throughout.

The intensities of the physically relevant actual and primary turbulences are readily measured by the
magnetic Reynolds number and the Lundquist number,
\begin{equation}
\Rm=\urms/\eta\kf,\quad\Lu=\brms/\eta\kf,
\end{equation}
where $\urms$ and $\brms$ are the rms values of fluctuating velocity and magnetic field, respectively,
and $\kf$ is the effective
forcing wavenumber.

\section{Results}
Throughout this section we set the mean flow $\meanUU$ to zero.	
An important criterion for the correctness of the generalized
test-field methods is the agreement of their
results with those of the imposed-field method.
In most cases we checked for this criterion.
Of course, the imposed-field method is only applicable if the actual
mean field in the main run is uniform.
If this is not the case, we are in some instances still able to
perform validation by comparing with analytical results.

\subsection{Zero mean magnetic field}
In this section we assume that the mean field is absent or
weak enough as not to affect 
the fluctuating fields markedly, that is, $\uu\approx\uuN$, $\bb\approx\bbN$.  
In particular it can then not render the transport coefficients  anisotropic.
Therefore, we denote by $\alpha$ and $\etat$ simply the average
of the first two diagonal
components of $\aTens$ and $\bTens$, i.e.\
$\alpha=(\alpha_{11}+\alpha_{22})/2$ and
$\etat=(\eta_{11}+\eta_{22})/2$, respectively.
If not specified otherwise we set
$\itilde{B}_{\text{imp}}=10^{-3}$ or zero.
Furthermore, we take $\Pm=1$, i.e.\ $\nu=\eta$.
 
\subsubsection{Purely hydrodynamic forcing}
\label{robForc}
In order to make contact with known results, we consider first the
case of the hydrodynamically driven Roberts flow.
In two dimensions, no small-scale dynamo is possible,
hence $\bbN=\zervec$ and $u_{0\rms}=N_{\rm K}/\nu\kf^2$.
In three dimensions, however, this solution could be unstable, but 
we have not yet employed sufficiently large $\Rm$ allowing for that.
For $\Rm\ll1$, $\alpha$ is given by (Brandenburg et al.\ 2008a)
\begin{equation}
\alpha/\alpha_{\rm0K}=\Rm/[1+(k_z/\kf)^2],\quad \alpha_{\rm0K}=-\urms/2,
\label{alpRoberts}
\end{equation}
where $k_z$ is the  wavenumber of the harmonic test fields. 
The minus sign in $\alpha_0$ accounts for the fact that the Roberts flow has positive
helicity, which results in a negative $\alpha$.

\begin{figure}[t!]\begin{center}
\includegraphics[width=\columnwidth]{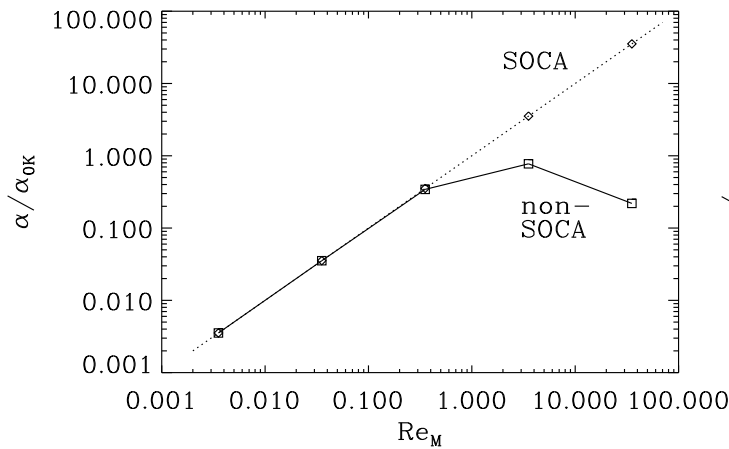}
\end{center}\caption[]{
$\alpha/\alpha_{\rm 0K}$ vs. $\Rm$ for purely kinetic Roberts forcing
with $\Pm=1$ and $k_z=0$ (2D case)
from the quasi-kinematic and all versions of the generalized method (solid line with squares).
Note the full agreement with Eq.~\eqref{alpRoberts} (dotted line)
for $\Rm\ll1$.
Diamonds indicate the results of the test-field methods with the
$\FFF_T'$ and $\EEE_T'$ terms in
\eqsref{duTdt}{daTdt} neglected, again coinciding with  \eqref{alpRoberts}.
}\label{pRobK_Fdep_standard}\end{figure}

Making use of the quasi-kinematic method,
as well as of all four versions of the generalized method,
we calculated $\alpha$ for $N_{\rm M}=0$,
$k_z=0$ (2D case) and values of $\tilde{N}_{\rm K}$ ranging from $0.01$ to $100$
with a ratio of 10; $\tilde{u}_\rms$ grows then from $0.005$ to $50$.
\FFig{pRobK_Fdep_standard}
shows $\alpha/\alpha_0$ versus $\Rm$ (solid line). 
Here the data points for all methods are
indistinguishable.
All results also agree with those of the imposed-field method.

Agreement with the SOCA result \eqref{alpRoberts} (dotted line) exists
for $\Rm\ll1$.
For $\Rm>1$, this is not applicable, because dropping the $\EEE_T'$ term
is then no longer justified.
The SOCA values are nevertheless numerically reproducible
by the test-field methods when ignoring
the $\FFF_T'$ and $\EEE_T'$ terms in \eqsref{duTdt}{daTdt};
see the diamond-shaped data points in  \Fig{pRobK_Fdep_standard}.

Corrections to the result \eqref{alpRoberts} with the $\EEE_T'$ term retained
were computed analytically by R\"adler et al.\ (2002a,b).
The corresponding values are again well reproduced by all flavors
of the generalized test-field method
as well as by the imposed-field method.

In the first line of \Tab{Talpha3D}, 
we repeat the result for  $\tilde{N}_{\rm K}=1$
and added that for test fields with the wavenumber $k_z=1$,
from where we also come to know the turbulent diffusivity $\etat$.
Note the difference between the $\alpha$ values for $k_z=1$ and $k_z=0$
roughly given by the factor $\sqrt{2}$ from \eqref{alpRoberts} for $k_z=k_f=1$.
Additionally, the results of the quasi-kinematic method for $k_z=1$, $\alphaQk$ and $\etatQk$, are shown.
As expected, they coincide completely with $\alpha$ and $\etat$.

\subsubsection{Purely magnetic forcing}
\begin{figure}[t!]\begin{center}
\includegraphics[width=\columnwidth]{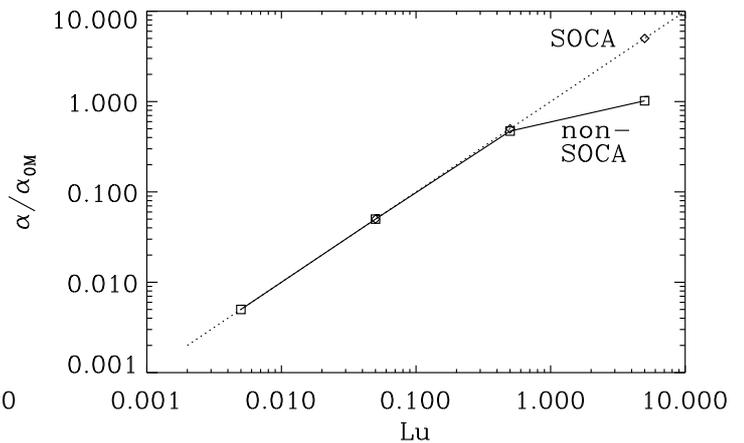}
\end{center}\caption[]{
$\alpha/\alpha_{\rm0M}$ versus $\Lu$ for purely magnetic Roberts forcing with $\Pm=1$ and $k_z=0$ (2D case)
from all versions of the generalized method (solid line with squares).
Note the full agreement with \Eq{alpRoberts1} (dotted line) for $\Lu\ll1$.
Diamonds 
give the results of the test-field methods with the $\EEE_T'$ and $\FFF_T'$ terms in
\eqref{FTandETstandard} neglected, again coinciding with  \eqref{alpRoberts1}.
}\label{pRobM_Edep}\end{figure}

Next we consider the case of purely magnetic Roberts forcing
i.e.\ $N_{\rm K}=0$.
Due to its Beltrami property, $\bbN \propto \ff$, $\uuN=\zervec$
is a solution of \eqsref{du0dt}{da0dt}.
A bifurcation leading to solutions with $\uuN\ne\zervec$ cannot be ruled out generally,
but was never observed.
Thus we have for the rms value of the
magnetic vector potential $a_{0\rms}=N_{\rm M}/\eta\kf^2$, hence 
$b_{0\rms}=N_{\rm M}/\eta\kf$.
The appropriate parameter for expressing the strength of the
fluctuating field(s) is now
the Lundquist number and the corresponding analytic result for $\Lu\ll1$ reads
\begin{equation}
\alpha/\alpha_{\rm0M}= (\Lu/\Pm)/[1+(k_z/\kf)^2],\quad \alpha_{\rm0M}=3\brms/4
\label{alpRoberts1}
\end{equation}
(for the derivation see \App{PurelyMagneticRoberts}).
It turns out that the sign of $\alpha$ coincides now
with that of the helicity of the forcing function.
Again, we consider first the two-dimensional case with $k_z=0$;
see \Fig{pRobM_Edep}.
In analogy to purely hydrodynamic forcing we find for $\Lu\ll1$
agreement between all versions of the generalized test-field method
(solid line with  squares) with \Eq{alpRoberts1} (dotted line).
For higher values the SOCA versions (see \sectref{robForc}) accomplish
the same; see diamond data points.
Note, that for the last data point with $\Lu=7$ it was necessary
to lower the strength of the imposed field to $\Bimp/\eta k_1=10^{-4}$,
because otherwise the solution of the main run becomes unstable
and changes to a new pattern.

\begin{table}[t]
\begin{threeparttable}[b]
\caption{
Kinematic results for $\tilde\alpha$ and $\tilde\eta_{\rm t}$
for purely hydrodynamic 
($\tilde{N}_{\rm M}=0$)
purely magnetic ($\tilde{N}_{\rm K}=0$),
and hydromagnetic Roberts forcing, $\Pm=1$.
The wavenumber of the test field is $k_z=1$,
except in the third column where $k_z=0$.
Its results agree with those of the imposed-field method.
$\tilde\alpha^\QK$ and $\tilde\eta_{\rm t}^\QK$ refer to the quasi-kinematic method.
}
\vspace{12pt}
\begin{tabular}{@{\hspace{4pt}}c@{\hspace{6pt}}c@{\hspace{8pt}}c@{\hspace{8pt}}c@{\hspace{8pt}}c@{\hspace{8pt}}c@{\hspace{8pt}}l@{\hspace{4pt}}}
$\tilde{N}_{\rm K}$ & $\tilde{N}_{\rm M}$ & $\tilde\alpha(k_z=0)$ 
& $\tilde\alpha$ & $\tilde\alpha^\QK$ 
& $\tilde\eta_{\rm t}$ & $\tilde\eta_{\rm t}^\QK$ \\
\hline
1           & 0 &  $-0.0857$    &$-0.0569$     & $-0.0569$                      & $0.0399$  &$0.0399$ \\
0           & 1 &  $\ivm0.2499$   &$\ivm0.1684$    & $\ivm0.0000$  & $0.1188$  &$0.0000$ \\[3pt]
3.364   & 0 &  $-0.7330$    &$-0.4734$     & $-0.4734$                      & $0.3087$  &$0.3087$ \\
%
0           & 1.942 &  $\ivm0.8219    $   &$\ivm0.5664$    & $\ivm0.0000$  & $0.3983$  &$0.0000$ \\
3.364   & 1.942 &  $-0.0081$   &$\ivm0.0664$    &                     $-0.4734$  & $0.6604$  &$0.3086$ \\[4pt] 
3.364   & 0 &  $-1.0002$    &$-0.6668$     & $-0.6666$  & $0.4715$  &$0.4714$ \tnote{1}\\
%
0          & 1.942 &  $\ivm1.0000$  & $\ivm0.6666$  & $\ivm0.0000$  & $0.4714$  &$0.0000$ \tnote{1}\\
3.364  & 1.942 &   $-4\!\cdot\!10^{-6}$&     $\ivm2\!\cdot\!10^{-5}$        & $-0.6666$  & $0.9428$  &$0.4714$  \tnote{1}\\
\label{Talpha3D}\end{tabular}
\begin{tablenotes}
\item [1] with SOCA
\end{tablenotes}
\end{threeparttable}
\end{table}

The second line of Table \ref{Talpha3D} repeats 
the result for $\tilde{N}_{\rm M}=1$, again
amended by those for $k_z=1$ and 
the results of the quasi-kinematic method which is obviously
unable to produce correct answers.
This is because
the mean electromotive force is now given by $\overline{\uuB\times\bbN}$. which is
only taken into account in the generalized method.
Note further that $\etat$ is positive both for hydrodynamic and
magnetic forcings.

\subsubsection{Hydromagnetic forcing}

\label{kinMHDforce}
As already pointed out in \sectref{forcing}, in the absence of a mean field,
for simultaneous kinetic and magnetic Roberts forcing
there is a solution of \eqsref{du0dt}{da0dt} consisting just of the solutions $\uuN$ and $\bbN$ of the system forced purely hydrodynamically
and magnetically, respectively. Again, a bifurcation leading to another type of solution cannot be ruled out, but was not observed.
Only within SOCA, however, the decoupling of $\uuN$ and $\bbN$
lets the value of $\alpha$ for hydromagnetic forcing be purely additive in the values for purely hydrodynamic and purely magnetic forcings. 
We denote the two latter
by $\aTens_\sk = \aTens(\bbN=\zervec)$ and $\aTens_\sm=\aTens(\uuN=\zervec)$, respectively.
When abandoning SOCA, the terms
$(\uuB\times\bbN)'$ and $(\jjN\times\bbB+\jjB\times\bbN)'$
in \Eqs{du1dt}{da1dt} provide couplings between $\uuB$ and $\bbB$
and give rise to an additional ``magnetokinetic" part, 
of $\alpha$ defined as $\aTens_\smk=\aTens - \aTens_\sk -\aTens_\sm$.
Note that we use here lower case subscripts k, m, and mk
to distinguish from the split introduced at the end of
\Sec{GeneralTreatment}, which applies to the nonlinear case.
In contrast, the occurrence of $\aTens_\smk$ is a purely kinematic effect.
While $\aTens_\sk$ and $\aTens_\sm$ are, to leading order (and hence in SOCA),
quadratic in the respective background fluctuations
the magnetokinetic term is of leading fourth order and is
representable in schematic form as
$\aTens_\smk\propto\langle \uuN \ppsiN\,\bbN \aaN\rangle$.

Lines 5 and 8 of \Tab{Talpha3D} show cases with hydromagnetic forcing
and amplitudes adjusted such that we would have
$\tilde\alpha_\sk =\tilde\alpha_\sm=1$ if SOCA were valid.
In either case the preceding two lines present the corresponding purely forced cases. Lines 6 to 8 refer to the SOCA version of the
test-field methods.
It can be clearly seen that the results are additive only in the latter case. The value of $\aTens_\smk$ as inferred from lines 3 to 5 is
$-0.1$ resulting in a considerable reduction of $\alpha$ in comparison with the purely additive value.
This is owed to the strong forcing amplitudes leaving the applicability ranges of SOCA far behind.

\begin{figure}[h]
\begin{center}
\includegraphics[width=\linewidth]{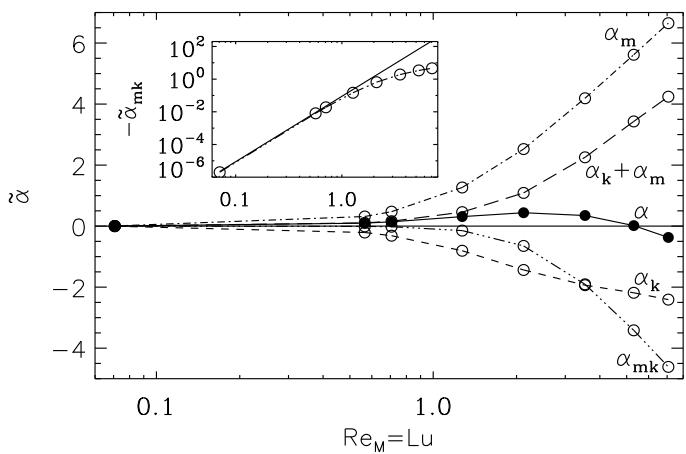}
\end{center}\caption[]{\label{pRobMK_ubrmsDep_alp}
$\alpha$ versus $\Rm=\Lu$ for hydromagnetic Roberts forcing with $\Pm=1$ and $k_z=0$ (2D case).
Along with the total value the constituents ,$\alpha_\sk$, $\alpha_\sm$ and $\alpha_\smk$ as well as $\alpha_\sk+\alpha_\sm$ are shown.
Note the sign change in $\alpha$ at $\Rm\approx0.54$.
The inset shows $\alpha_\smk$ in comparison to the result of a fourth order
analytical calculation (solid line). }
\end{figure}

Figure \ref{pRobMK_ubrmsDep_alp} shows $\alpha_\smk$ for equally strong velocity and magnetic fluctuations as a function of $\Rm=\Lu$
together with $\alpha_\sk$, $\alpha_\sm$, $\alpha_\sk+\alpha_\sm$ and the resulting total value $\alpha$.
Note the significant difference between  the naive extrapolation of SOCA, $\alpha_\sm+\alpha_\sk$, and the true $\alpha$.
In its inset the figure shows the numerical values of $\alpha_\smk$ in comparison to the result of a fourth order calculation
$\alpha_\smk = - \sqrt{2}/64\urms^2\brms^2$ (for the derivation see \App{alpmkDer}). 
Clearly, the validity range of this expression extends beyond $\Rm=\Lu=1$ and hence further
than the one of SOCA.
It remains to be studied whether the magnetokinetic contribution
has a significant effect also in the more general case when
$\uuN\not\parallel\bbN$.
If so, considering $\alpha$ to be the sum of a kinetic and a magnetic part,
as often done in quenching considerations, may turn out to be too simplistic.
\begin{figure*}[t]
\begin{center}
\includegraphics[width=\textwidth]{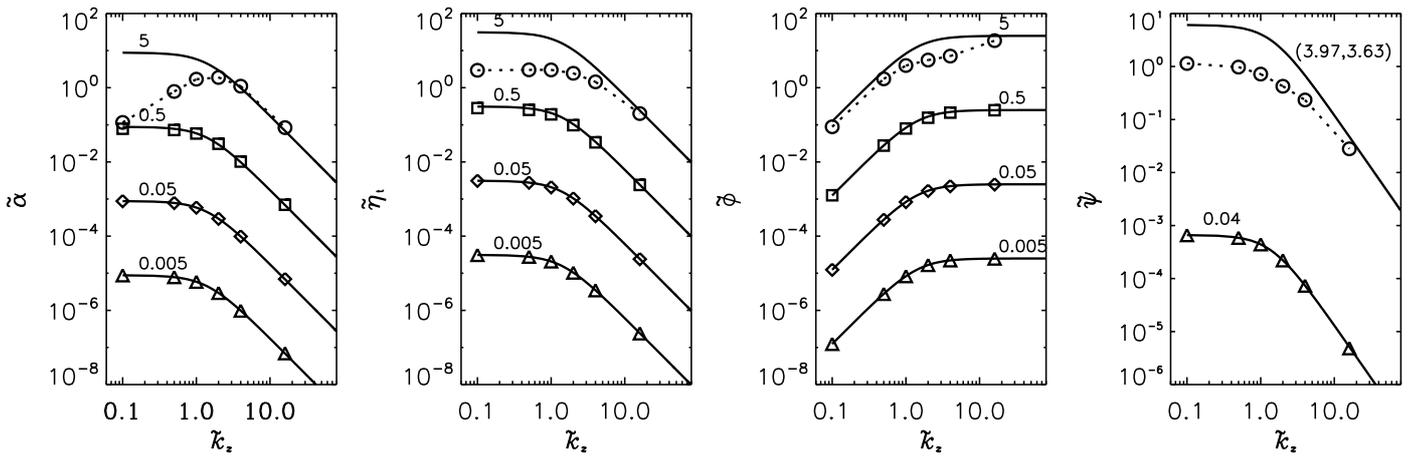}
\end{center}\caption[]{\label{pRobMK_k1}
$\alpha(k_z)$, $\etat(k_z)$, and $\phi(k_z)$ for
hydromagnetic Roberts forcing with $\Pm=1$ and $\sigma=1$ (left three panels), likewise  $\psi(k_z)$, but for $\sigma=0.5$ (rightmost panel).
Solid lines correspond to SOCA results, cf.\ \App{phideriv}.
Curve labels refer  to $\Rm=\Lu$ or $(\Rm,\Lu)$.
}
\end{figure*}
Likewise one may wonder whether closure approaches to the determination of
transport coefficients supposed to be superior to SOCA can be successful
at all as long as they do not take fourth order correlations into
account properly.

\paragraph{For the tensors $\phiTens$ and $\psiTens$,} which turn out to
show up with simultaneous hydromagnetic and magnetic forcing
only (in addition, $\phiTens$ requires $z$-dependent mean fields)
we have of course again isotropy,
$\phi_{11}=\phi_{22}\equiv\phi$, $\psi_{11}=\psi_{22}\equiv\psi$.

As a peculiarity of the Roberts flow, $\psi$ vanishes
in the range of validity of SOCA
if the helicity is maximum ($\sigma=1$ in \eqref{robForce}).
For this case the
first three panels of \Fig{pRobMK_k1} show the dependences $\alpha(k_z)$, $\eta(k_z)$ and $\phi(k_z)$ with
different values of $u_{0\rms}=b_{0\rms}$
(data points, dotted lines).
The last panel shows $\psi(k_z)$ for $\sigma=0.5$ and the same forcing amplitudes as before.
As explained above, $\uuN$ and $\bbN$ can now no longer be forced independently from each other.
Hence both fields can not show exactly the geometry defined by \eqref{robForce} and
$\urms$ and $\brms$ diverge increasingly with increasing forcing.

As demonstrated in \App{phideriv},
$\phi(k_z)\propto k_z^2/( k_z^2+\kf^2)$, 
$\alpha(k_z),\eta(k_z),\psi(k_z)\propto 1/( k_z^2+\kf^2)$
in the SOCA limit.
For comparison these functions are depicted by solid lines. 
Note the clear deviations from SOCA for $\Rm=\Lu=5$, particularly in $\alpha$.
Note also that the expression for $\psi$ was derived under the assumption that the background has the geometry \eqref{robForce}.
It is therefore not applicable in a strict sense. The clear disagreement with the values of $\psi$ from the test-field method
for the high values of $\Rm$ and $\Lu$ are hence not only due to violating the validity constraint $\Rm\ll1$.

\subsection{Dependence on the mean field}
We now admit dynamically effective mean fields
and hence have to deal with anisotropic fluctuating fields
$\uu$ and $\bb$
which result in anisotropic tensors $\aTens$, $\bTens$, $\phiTens$ and $\psiTens$.
For the chosen forcing, $\meanBB$ is the only source of anisotropy in the $xy$ plane, so
$\aTens$  has to have the form 
\[
    \alpha_{ij} = \alpha_1 \delta_{ij}  + \alpha_2 \hatB_i \hatB_j,\quad i,j=1,2,
\]
with $\hatBB$ the unit vector in the direction of $\meanBB$ (here the $x$ direction).
We obtain then
$\alpha_{11}=\alpha_1+\alpha_2$ and $\alpha_{22}=\alpha_1$.
Of course, the tensors $\bTens$, $\phiTens$ and $\psiTens$ are built analogously.
Clearly, irrespective whether the forcing is pure or mixed
the effects of $\Bimp$ prevent the fluctuating $\uu$ and $\bb$ from  having Roberts geometry .

For vanishing magnetic background,
$\bbN=\zervec$, the generalized methods are still expected
to give results coinciding with those 
of the quasi-kinematic one, but with
$\bbN\neq\zervec$
we will leave safe mathematical grounds and enter empirical work.

\subsubsection{Purely hydrodynamic forcing}  
\label{BdepM}

\begin{figure}[t!]\begin{center}
\includegraphics[width=\columnwidth]{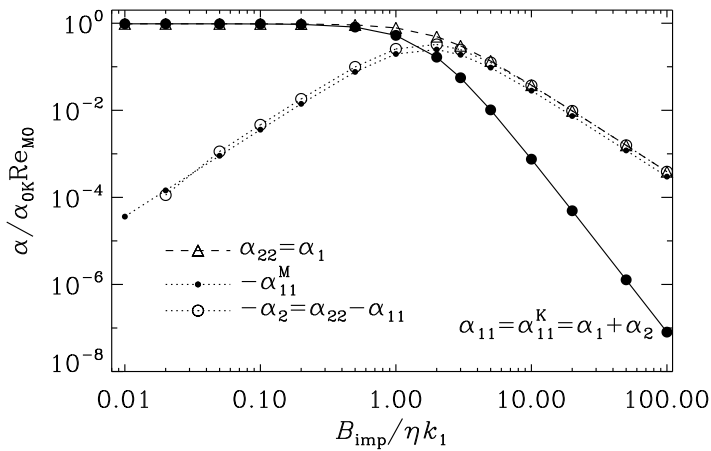}
\end{center}\caption[]{
$\alpha_{11}$ (solid line, filled circles) and
$\alpha_{22}$ (dashed line, open triangles)
as functions of the imposed field strength $\Bimp$,
compared with $-\alpha_{11}^\sM$ (dotted line, small dots) and 
$\alpha_{22}-\alpha_{11}=-\alpha_{2}$ (dotted line, open circles)
for purely kinetic Roberts forcing with $\tilde{N}_{\rm K}=1$ and $\Pm=1$.
$\alpha_{11}^\sM\approx\alpha_2$ throughout.
Note that $\alpha_{0\sK}<0$ and that the $\alpha$ symbols in the legend refer
to the normalized (hence sign-inverted) quantities.
}\label{pRobK_Bdep}\end{figure}

In this case we have $\EEEEB = \overline{\uu\times\bbB}=\EEEEBK$ and all flavors of the generalized method have to 
yield results which coincide with those of the quasi-kinematic method.
This is valid to very high accuracy for the {\sf ju} and {\sf bu}
versions and somewhat less perfectly so for the {\sf bb} and {\sf jb} versions.
We emphasize that the presence of $\Bimp$, although being solely
responsible for the occurrence of magnetic fluctuations, does not result
in a failure of the quasi-kinematic method as one might conclude from
the model used by Courvoisier et al.\ (2010).

Figure \ref{pRobK_Bdep} presents 
the constituents of $\aTens$ as functions
of the imposed field in the 2D case with $\Pm=1$
We may conclude from the data
that $\alpha_2$ is negative and approximately equal to $\alpha_{11}^M$.
For values of $\Bimp/\eta k_1> 5$ its modulus
approaches $\alpha_{22}=\alpha_1$ and thus gives rise to the strong quenching of the effective $\alpha=\alpha_{11}$.
Indeed, $\alpha(\Bimp)$ can be described by a power law with an exponent $-4$ for large $\Bimp$.
By comparing with computations in which the non-SOCA term was neglected,
we have checked that this discrepancy is not a consequence of SOCA.
This is at odds with analytic results predicting either $\alpha\propto B^{-2}$
(Field et al.\ 1999; Rogachevskii \& Kleeorin 2000) or $\propto B^{-3}$
(Moffatt 1972; R\"udiger 1974).
Sur et al.\ (2007) suggested that this difference was due to the fact
that the flows are either time-dependent or steady.
However, inspecting their Figure~2, their numerical values for $\alpha$
do exhibit the $B^{-4}$ power law.
They also found that a $\alpha^{\rm M}$,
defined similarly to our $\alpha_{11}$,
increases quadratically with $\meanBB$ for weak fields and
declines quadratically for strong fields (Sur et al.\ 2007).
This is in turn in agreement with our present results.

\subsubsection{Purely magnetic forcing}  
Here, the mean electromotive force is simply
$\meanEEEE= \EEEEBM =\overline{\uuB\times\bb}$.
This is true as long as
significant velocities in the main run occur only due to
the presence of a mean field, that is, as long as $\uuN=\zervec$ (see above). 
While $\meanBB$ is weak, $\meanEEEE$ is approximately $\overline{\uuB\times\bbN}$.
However, one could speculate that, if the imposed field reaches
appreciable levels, i.e., if $\uu$ is sufficiently strong,
$\meanEEEE$ can, with good accuracy, be approximated by
$\EEEEBK=\overline{\uu\times\bbB}$.
Since the quasi-kinematic method takes just this term into account,
it should then produce useful results.

\begin{figure}[t!]\begin{center}
\includegraphics[width=\columnwidth]{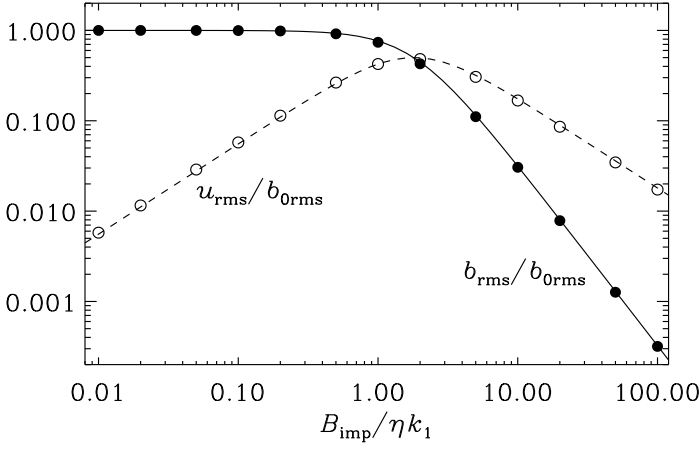}
\end{center}\caption[]{
Root-mean-square values $\brms$ (filled circles) and $\urms$ (open circles) as functions
of the imposed field strength $\Bimp$
for purely magnetic Roberts forcing $\tilde{N}_{\rm M}=1$ with $\Pm=1$.
The solid and dashed lines represent the fits given by \Eq{BrmsUrmsfit}.
}\label{pRobM_Bdep_ubrms}\end{figure}

In \Fig{pRobM_Bdep_ubrms} we show the rms values of the resulting magnetic
and velocity fields as functions of the imposed field strength for
$\tilde{N}_{\rm M}=1$, corresponding to $\Lu=1/2$ if $\BBimp=\zervec$.
The data points can be fitted by expressions of the form
\begin{equation}
{\brms\over b_{0\rms}}={1\over1+\Bimp^2/B_*^2},\quad
{\urms\over b_{0\rms}}={\Bimp/B_*\over1+\Bimp^2/B_*^2},
\label{BrmsUrmsfit}
\end{equation}
where $B_*\approx1.8\,\itilde{N}_{\rm M}$.
Note, that indeed the velocity fluctuations become dominant over the magnetic
ones for $\Bimp/\eta k_1>2$.

\begin{figure}[t!]\begin{center}
\includegraphics[width=\columnwidth]{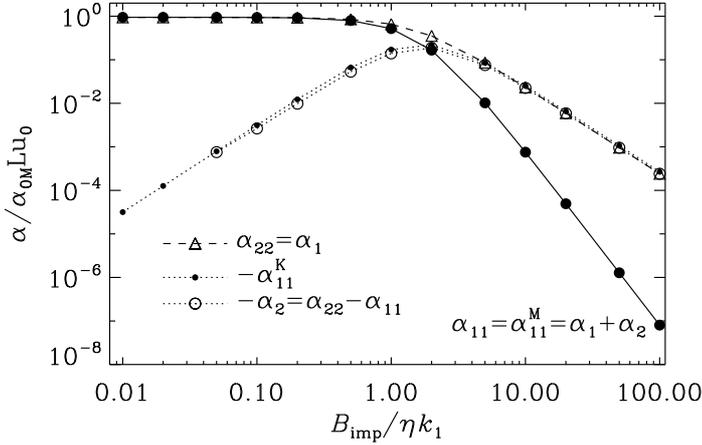}
\end{center}\caption[]{
$\alpha_{11}$ (solid line, filled circles) and
$\alpha_{22}$ (dashed line, open triangles)
as functions of the imposed field strength $\Bimp$,
compared with $-\alpha_{11}^\sK$ (dotted line, small dots) and 
$\alpha_{22}-\alpha_{11}=-\alpha_{2}$ (dotted line, open circles)
for purely magnetic Roberts forcing with $\tilde{N}_{\rm M}=1$ and $\Pm=1$.
Note that $\alpha_{11}^\sK\approx\alpha_2$ throughout.
}\label{pRobM_Bdep}\end{figure}

\begin{figure}[t!]\begin{center}
\includegraphics[width=\columnwidth]{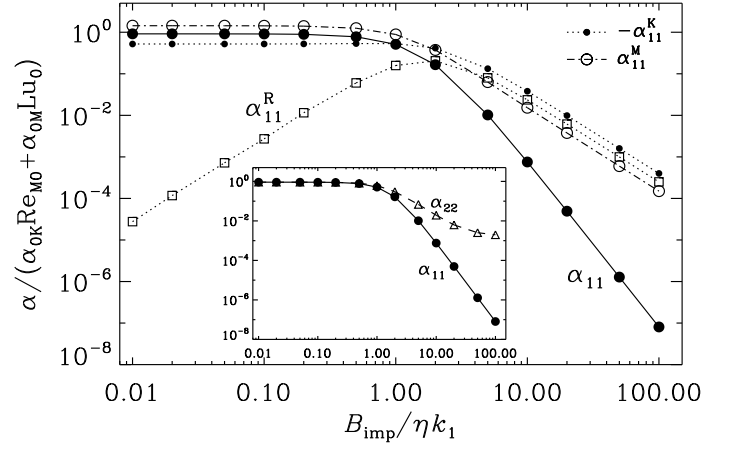}
\end{center}\caption[]{
$\alpha_{11}$ (solid line, filled circles)
as function of the imposed field strength $\Bimp$,
compared with $-\alpha_{11}^\sK$ (dotted line, small dots), 
$\alpha_{11}^\sM$ (dash-dotted line, open circles) and
$\alpha_{11}^{\rm R}$ (dotted line, open squares)
for hydromagnetic Roberts forcing with $\tilde{N}_{\rm M}=\tilde{N}_{\rm K}=1$ and $\Pm=1$.
The inset shows $\alpha_{22}$ (dashed line, open triangles)
compared to $\alpha_{11}$.
}\label{pRobMK_Bdep}\end{figure}

The resulting finding, as shown in \Fig{pRobM_Bdep}, is completely
analogous to the one of \sectref{BdepM}, but now we see
$-\alpha_{11}^\sK \approx -\alpha_2$ approaching $\alpha_{22}=\alpha_1$
with increasing $\Bimp$.
Hence, the supposition that the quasi-kinematic method could give
reasonable results for strong mean fields has not proven true as
$\alpha_{11}^\sK$ is not approaching $\alpha_{11}$, despite the domination
of $\urms$ over $\brms$.
Instead, the values from the quasi-kinematic method have the wrong sign and
deviate in their moduli by several orders of magnitude.

\begin{table}[b!]\caption{
Dependence of the diagonal components of the $\alpha$ tensor
on $\itilde{B}_{\rm imp}$ for $\itilde{N}_{\rm M}=1$, $\Pm=1$ using the
generalized method ($\tilde\alpha_{11}$ and $\tilde\alpha_{22}$)
together with the kinetic contribution
$\tilde\alpha_{11}^\sK$ and the results from
the quasi-kinematic method ($\tilde\alpha_{11}^{\QK}$ and
$\tilde\alpha_{22}^{\QK}$).
}\vspace{12pt}\centerline{\begin{tabular}{lrrrr}
\multicolumn{1}{c}{${\tilde B}_{\rm imp}$}
& \multicolumn{1}{c}{$10^{-2}$}
& \multicolumn{1}{c}{$1$}
& \multicolumn{1}{c}{$10^1$}
& \multicolumn{1}{c}{$10^2$}\\
\hline
$\tilde\alpha_{11}$           & $2.499\;10^{-1}$      & $1.376\;10^{-1}$      & $2.000\;10^{-4}$       & $2.131\;10^{-8}$ \\
$\tilde\alpha_{22}$           & $2.499\;10^{-1}$      & $1.747\;10^{-1}$      & $6.161\;10^{-3}$       & $6.390\;10^{-5}$ \\
$\tilde\alpha_{11}^\sK$   &$\!\!\!\!-8.391\;10^{-6}$&$\!\!\!\!-4.540\;10^{-2}$& $\!\!\!\!-6.666\;10^{-3}$&$\!\!\!\!-7.067\;10^{-5}$\\
$\tilde\alpha_{11}^{\QK}$ &$\!\!\!\!-7.858\;10^{-6}$&$\!\!\!\!-4.350\;10^{-2}$& $\!\!\!\!-6.657\;10^{-3}$&$\!\!\!\!-7.067\;10^{-5}$\\
$\tilde\alpha_{22}^{\QK}$ &$\!\!\!\!-2.247\;10^{-7}$&$\!\!\!\!-1.152\;10^{-3}$& $\!\!\!\!-4.740\;10^{-7}$&$\!\!\!\!-5.326\;10^{-13}\!\!$\\
\label{Talpha}\end{tabular}}\end{table}

In \Tab{Talpha} we compare, for different values of $\Bimp$,
the values of $\alpha_{11}$ and $\alpha_{22}$,
obtained using the generalized test-field method,
with those of $\alpha_{11}^\sK$
and those from the quasi-kinematic method,
$\alpha_{11}^{\QK}$ and $\alpha_{22}^{\QK}$,
where the entire dynamics of $\bbB$ has been ignored.
Note, again, that the results of all four version of the generalized
test-field method agree with each other.

\subsection{Hydromagnetic forcing}

\begin{figure}[t!]\begin{center}
\includegraphics[width=\columnwidth]{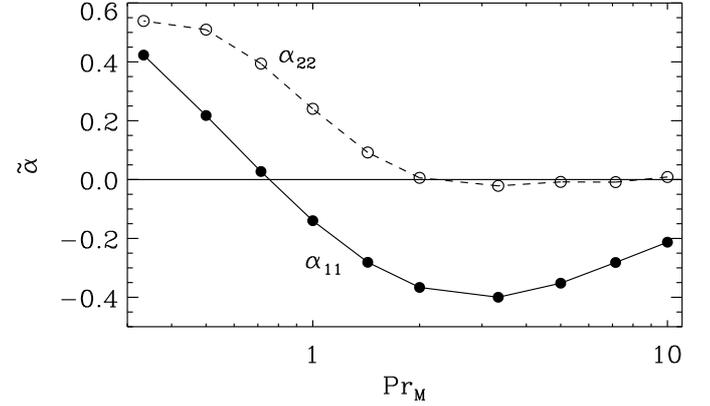}
\end{center}\caption[]{
Dependence of $\alpha_{11}$ and $\alpha_{22}$ on $\Pm$
for hydromagnetic Roberts forcing with $\Lu/\Pm=1$ and $\Bimp/\nu k_1=1$.
}\label{pRobMixed_PrM}\end{figure}

Given that the $\alpha$ effect can be sensitive to the value of $\Pm$,
we study $\alpha_{11}$ and $\alpha_{22}$ as functions of $\Pm$, keeping
$\Lu/\Rm=1$ and $\Bimp/\nu k_1=1$ fixed.
The result is shown in \Fig{pRobMixed_PrM}.

In the interval $2\le\Pm\le10$ the $\alpha$ coefficients exhibit a high sensitivity with respect to $\Pm$ changing even
their sign at $\Pm\approx 0.7$ and $2$, respectively. Note also the occurrence of a minimum
and the subsequent growth for $\Pm>4$.

Analogously to \Figs{pRobK_Bdep}{pRobM_Bdep} we show in Figure \ref{pRobMK_Bdep} the constituents of $\aTens$ versus $\Bimp$.
Note that we have used here $\alpha_{\rm0K}\Rey_{\rm M0}+\alpha_{\rm0M}\Lu_0>0$ for normalizing $\aTens$ which is the kinematic value 
of $\alpha_{11}=\alpha_{22}$ for $k_z=0$ and small $u_{0\rms}$, $b_{0\rms}$; see \Eqs{alpRoberts}{alpRoberts1}, \sectref{kinMHDforce}.

It can be observed that $\alpha_{11}^\sM$ at first dominates over $-\alpha_{11}^\sK$, but at $\Bimp/\eta k_1 \approx 10$
their relation reverses. Remarkably, the ratio of the moduli reaches, for high values of $\Bimp$, just the inverse of that 
for low values.
The strong quenching of $\alpha_{11}$ is now a consequence of $\alpha_{11}^{\rm R}$ approaching $-\alpha_{11}^\sK-\alpha_{11}^\sM$. 
In complete agreement with the former two cases with pure forcings,
$-\alpha_{11}$ is proportional to $\Bimp^{-4}$.
However, we see a deviating behavior of $\alpha_{22}(\Bimp)$
as it is no longer following a power law.

\subsection{Convergence}

In most of the cases the four different versions of the generalized method,
(see \Tab{TableVersions})
give quite similar results.
For purely hydrodynamic and purely magnetic forcing
there is agreement to all significant digits.
The agreement becomes somewhat less certain when there
is hydromagnetic forcing, i.e.\
$N_{\rm K}\neq0$, $N_{\rm M}\neq0$.
In general, however, agreement is improved by increasing the numerical resolution.

\begin{figure}[t!]\begin{center}
\includegraphics[width=\columnwidth]{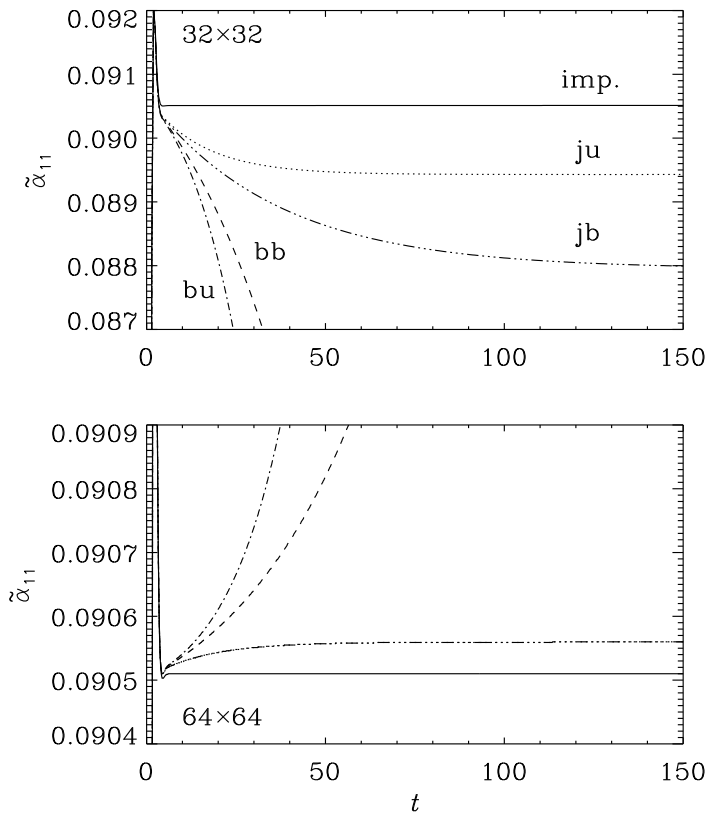}
\end{center}\caption[]{
Convergence of $\alpha_{11}$
the {\sf ju} and {\sf jb} versions of the generalized method
with the result of the imposed-field method and exponential
divergence of the versions {\sf bu} and {\sf bb}
for $\Pm=1$ with
$\itilde{N}_\sK=\itilde{N}_\sM=1$,
$\tilde{B}_0=1$, $k_z=0$ and a resolution of either $32^2$ mesh points
(upper panel) or $64^2$ mesh points (lower panel).
Note the significant improvement of the converging methods' agreement by 
doubling the resolution: The deviation is changing from $\approx 2.5$\% to $\approx 0.05$\%,
that is, by a factor close to $2^6$ suggested by the sixth order
of the difference scheme.
}\label{pncomp}\end{figure}

Yet another complication arises when $B_0\neq0$, because then some of
the versions display exponentially growing solutions; see \Fig{pncomp}.
We see no other explanation than that by the deliberate rearrangements 
leading to \eqref{FTd}, \eqref{ETd}
a potential for spurious instabilities was introduced.

Their real occurrence, however, depends obviously 
on intricate properties of the fluctuating fields from the main run $\uu$ and $\bb$.
As the test equations are linear, unlimited exponential growth results
(at least for stationary fields in the main runs).
We suppose that if one could remove the unstable eigenvalues of the homogeneous version of the system
\eqref{duTdt}--\eqref{ETd} arbitrarily from its spectrum the solution of the inhomogeneous system would just be the correct one.

\section{Discussion}

The main purpose of the developed method consists in dealing with situations
in which hydrodynamic and magnetic fluctuations coexist.
The quasi-kinematic method can only afford those constituents of the
mean-field coefficients that are related solely to the hydrodynamic
background $\uuN$, but the new method is capable of delivering, in
addition, those related to the magnetic background $\bbN$.
Moreover, it is able to detect    
mean-field effects that depend on cross correlations of $\uuN$ and $\bbN$.
We have demonstrated this with the two fluctuations being
forced externally to have
the same Roberts-like geometry. With respect to $\alpha$
we observe a ``magneto-kinetic" part that is, to leading order, quadratic
in the magnetic Reynolds and Lundquist numbers and is capable of reducing
the total $\alpha$ significantly in comparison
with the sum of the $\alpha$ values resulting from purely hydrodynamic and purely magnetic backgrounds.
The tensors $\phiTens$ and $\psiTens$ which give rise
to the occurrence of mean forces proportional to
$\meanBB$ (or $\nab^2\meanBB$) and $\meanJJ$ are,
to leading order, bilinear in $\Rm$ and $\Lu$.
In nature, however,
external electromotive forces imprinting finite cross-correlations of $\uuN$
and $\bbN$ are rarely found.
Therefore the question regarding the
astrophysical relevance of these effects has to be posed.
Given the high values of $\Rm$ in practically all cosmic bodies,
small-scale dynamos are supposed to be ubiquitous and do indeed provide
hydromagnetic background turbulence.
But is it realistic to expect non-vanishing cross-correlations
under these circumstances?

Let us consider a number of similar, yet not completely identical
turbulence cells arranged in a more or less regular pattern.
We assume further that there is some asymmetry between upwellings and
downdrafts such that, say, the downdrafts are more efficient in amplifying magnetic
fields than the upwellings (Nordlund et al.\ 1992; Brandenburg et al.\ 1996).
As dynamo fields are solutions of the homogeneous induction
equation and the Lorentz force is quadratic in $\BB$,
bilinear cross-correlations,
$\overline{u_{0i}b_{0j}}$, obtained by averaging over
single cells can be expected to change their sign
randomly from cell to cell provided
the cellular dynamos have evolved independently from each other.
Consequently, the average over many cells would approach zero and the
aforementioned effects would not occur.
In contrast, cross-correlations that are even functions of the components
of $\bbN$ and their derivatives, were not rendered zero due to polarity
changes in the dynamo fields (e.g.\ the magneto-kinetic $\alpha$).

However, the assumption of independently acting cellular dynamos can
be put in question when the whole process beginning with the onset of
the turbulence-creating instability (e.g.\ convection)
is taken into account.
During its early stages, i.e.\ for small magnetic Reynolds numbers, the flow is at first
unable to allow for any dynamo action, but with growing amplitude
the large-scale dynamo can be excited first to create a
field that is coherent over many turbulence cells. With further growth
of its amplitude the (hydrodynamic) turbulence
eventually enters a stage in which small-scale dynamo action becomes possible.
The seed fields for these dynamos are now prevailingly  determined by the
mean field and due to its spatial coherence the polarity of the small-scale
field is not settling independently from cell to cell, thus potentially
allowing for non-vanishing cross-correlations.

But even if one wants to abstain from employing the influence of a
pre-existing mean field it has to be considered that neighboring cells
are never exactly equal. Thus, in the course of the growing amplitude
of the hydrodynamic background in some of them, the small-scale dynamo
will start working first hence setting the seed field for
its immediate neighbors. 
It is well conceivable that the field polarity initiated by one of the early starting
cells ``cascades" to more and more distant neighbors until this process
is limited by the cascades originating from other early starting cells.
The result could resemble the domains with uniform field orientation in ferromagnetic materials.
Consequently, we arrive at a
situation similar to the one discussed before, yet with less extended
regions of coinciding field polarity.
Instead of employing the idea
of a pre-existing large scale dynamo in the cosmic object at hand
one may even suppose that, given the smallness of the turbulence cells
compared to the scale of the embedding surroundings, there is
always a large-scale field, e.g.\ the galactic one,
that is coherent across a large number of turbulence cells.

In summary, cross-correlations and the mean-field effects connected to them are to be considered realistic options.
Direct numerical simulations employing the scenarios discussed above should be performed in order to clarify the significance
of these effects. This is equally valid for the cross-correlations related effects leading to $\meanEEEE_0$ (see Eq.\ \eqref{E0ansatz}).

\vspace{2mm}
In a recent paper, Courvoisier et al.\ (2010) discuss the range of applicability of the quasi-kinematic test-field method.
Their model consists of the equations of incompressible magneto-hydrodynamics with purely hydrodynamic forcing.
However, by imposing an additional uniform magnetic field $\boldsymbol{\cal B}$, together with the forced fluctuating velocity
a fluctuating magnetic field arises. It must be stressed that, following the further lines of their arguments, these 
fluctuations have to be considered as part of the {\em background} $(\uuN,\bbN)$, that is, representing just
those fluctuations that occur in the {\em absence} of the mean field.
This follows from the fact that, when defining transport coefficients such
as $\aTens$, the field $\boldsymbol{\cal B}$ is not regarded as part
of the mean field $\langle \BB \rangle$, in contrast to our treatment;
see their section 2.(b).
For simplicity they consider only the kinematic case and restrict the analysis to mean fields $\propto \ei{k_z z}$ with $k_z \rightarrow 0$.
In their main conclusion, drawn under these conditions, they state that the quasi-kinematic test-field method which considers only 
the magnetic response to a mean magnetic field must fail for  $\boldsymbol{\cal B}\ne\zervec$, that is $\bbN\ne\zervec$.
We fully agree in this respect, but should point out 
that the method was never claimed to be applicable in that case; see Brandenburg et al.\ (2008c, Sect.\ 3),
where it is mentioned that ``as in almost all supercritical runs a small-scale dynamo is operative,
our results which are derived under the assumption of its influence being negligible
may contain a systematic error.''
However, Courvoisier et al.\ (2010) overinterpret their finding in postulating
that already the determination of quenched coefficients such as
$\alpha(\meanB)$ for $\bbN=\zervec$ by means of the quasi-kinematic
method leads to wrong results.
The paper of Tilgner \& Brandenburg (2008), quoted by them
in this context, is just proving the correctness of the method,
as do Brandenburg et al.\ (2008c). 

Our tensor $\psiTens$ is related to their newly introduced
mean-field coefficient $\boldsymbol{\Gamma}$  by
$\psi_{ij}=\epsilon_{kj3}\Gamma_{i3k}$.
Unfortunately, an attempt to reproduce their results for
$\boldsymbol{\Gamma}$ (and likewise for $\aTens$) is not currently
possible owing to our modified hydrodynamics.
We postpone this task to a future paper.
 
\section{Conclusions}

Having been applied to situations with a magnetohydrodynamic background where both $\uuN$ and $\bbN$
have Roberts geometry, the proposed method has proven its potential 
for determining turbulent transport coefficients.
In particular, effects connected with cross-correlations between $\uuN$ and $\bbN$
could be identified and are in full agreement with analytical predictions as far as available.
No basic restrictions with respect to the magnetic Reynolds number
or the strength of the mean field in the main run, which causes the nonlinearity of the problem, are observed so far.
As a next step, of course, the simplifications in the hydrodynamics we used will be dropped, thus 
allowing to produce more relevant results and facilitating comparisons with 
work already done.
  
Due to the fact that we have no strict mathematical proof for its correctness,
there can be no full certainty about the general reliability of the method.
As a hopeful indication, in many cases, all four flavors of the method
produce practically identical results, but occasionally some of them show,
for unknown reasons, unstable behavior in the test solutions.
Clearly, further exploration of the method's degree of reliance by including three-dimensional and time-dependent
backgrounds is necessary. Homogeneity should be abandoned and backgrounds which come closer to real turbulence
such as forced turbulence or turbulent convection in a layer are to be taken into account.

Thus, the utilized approach of establishing a test-field procedure
in a situation where the governing equations are inherently nonlinear,
although by virtue of the Lorentz force only,
has proven to be promising.
This fact encourages us to develop
test-field methods for determining
turbulent transport coefficients connected with
similar nonlinearities
in the momentum equation.
An interesting target is the turbulent kinematic viscosity tensor,
and especially its off-diagonal components that can give rise to a mean-field   
vorticity dynamo (Elperin et al.\ 2007; K\"apyl\"a et al.\ 2009),
as well as the so-called anisotropic kinematic $\alpha$ effect
(Frisch et al.\ 1987; Sulem et al.\ 1989; Brandenburg \& von Rekowski 2001;
Courvoisier et al.\ 2010) and the $\Lambda$ effect (R\"udiger 1980, 1982).
Yet another example is given by the turbulent transport coefficients
describing effective magnetic pressure and tension forces due to the
quadratic dependence of the total Reynolds stress tensor on the mean
magnetic field (e.g., Rogachevskii \& Kleeorin 2007;
Brandenburg et al.\ 2010).

\appendix
\section{Incompressibility}
\label{IsothermalProblem}

The equations used in this paper had the advantage of simplifying the
derivation of the generalized test-field method, but the resulting flows
are not realistic because the pressure and advective terms are absent.
Here we drop these restrictions and derive
the test equations in the incompressible case with constant density.
The full momentum and induction equations take then the form
\begin{align}
{\partial\UU\over\partial t}
&=\UU\times\WW+\JJ\times\BB+\FF_{\rm K}+\nu\nabla^2\UU
-\nab P, \\
{\partial\AAA\over\partial t}&=\UU\times\BB+\FF_{\rm M}+\eta\nabla^2\AAA.
\end{align}
where $\WW=\curl\UU$ is the vorticity and
$P$ is the sum of gas and dynamical pressure
and absorbs the constant density.
The corresponding mean-field equations are
\begin{align}
{\partial\meanUU\over\partial t}
&=\meanUU\times\meanWW+\meanJJ\times\meanBB+\meanFFFF+\nu\nabla^2\meanUU
-\nab\meanP,\\
{\partial\meanAA\over\partial t}&=\meanUU\times\meanBB+\meanEEEE+\eta\nabla^2\meanAA,
\end{align}
where $\meanFFFF=\overline{\uu\times\ww}+\overline{\jj\times\bb}$
and $\meanEEEE=\overline{\uu\times\bb}$,
and the forcings were assumed to vanish on averaging.
The equations for the fluctuations are consequently
\begin{alignat}{2}
{\partial\uu\over\partial t}
&=\,&\meanUU\times\ww&+\uu\times\meanWW
+\meanJJ\times\bb+\jj\times\meanBB
\nonumber \\
&\phantom{=}&&+\FFF'+\FF_{\rm K}+\nu\nabla^2\uu-\nab p,
\label{dudt_h}\\
{\partial\aaaa\over\partial t}
&=&\!\meanUU\times\bb&+\uu\times\meanBB
+\EEE'+\FF_{\rm M}+\eta\nabla^2\aaaa,
\label{dadt_h}
\end{alignat}
where $\FFF'=(\uu\times\ww+\jj\times\bb)'$ and $\EEE'=(\uu\times\bb)'$.
As above we split the fields and likewise the
\Eqs{dudt_h}{dadt_h} into two parts, i.e.\ we write
$\uu=\uuN+\uuB$ and $\aaaa=\aaN+\aaB$
and arrive at
\begin{align}
{\partial\uuN\over\partial t}
&=\meanUU\times\wwN+\uuN\times\meanWW+\FFFN'+\FF_{\rm K}+\nu\nabla^2\uuN-\nab\pN,
\label{du0dt_h}\\
{\partial\aaN\over\partial t}
&=\meanUU\times\bbN+\EEEN'+\FF_{\rm M}+\eta\nabla^2\aaN,
\label{da0dt_h}
\end{align}
and the equations for the $\meanBB$ dependent parts
\begin{alignat}{2}
{\partial\uuB\over\partial t}
&=\,&\meanUU\times\wwB&+\uuB\times\meanWW
+\meanJJ\times\bb+\jj\times\meanBB
\nonumber \\
&&&+\FFFB'+\nu\nabla^2\uuB
-\nab\pB
\label{du1dt_h}\\
{\partial\aaB\over\partial t}
&=&\meanUU\times\bbB&+\uu\times\meanBB
+\EEEB'+\eta\nabla^2\aaB,
\label{da1dt_h}
\end{alignat}
where $\FFF'=\FFFN'+\FFFB'$ and $\EEE'=\EEEN'+\EEEB'$ with
$\FFFN'=(\uuN\times\wwN+\jjN\times\bbN)'$,
$\EEEN'=(\uuN\times\bbN)'$, and
\begin{align}
\FFFB'&=(\jjN\times\bbB+\jjB\times\bbN+\jjB\times\bbB
\nonumber \\
&\phantom{=}+\uuN\times\wwB+\uuB\times\wwN+\uuB\times\wwB)',\\
\EEEB'&=(\uuN\times\bbB+\uuB\times\bbN+\uuB\times\bbB)'.
\end{align}
We can rewrite these 
equations such that they become formally linear
in $\uuB$ and $\bbB$.
Following the pattern utilized in \sectref{GeneralTreatment} we find already for $\FFFB'$ four different ways of doing that.
Together with the two variants in the case of $\EEEB'$ we finally obtain eight flavors of the test-field method
where again in either case $\meanFFFF$ and $\meanEEEE$ are to be constructed analogously to $\FFFB'$ and $\EEEB'$.
One of these flavors is defined by
\begin{align}
\FFFB'&=(\uu\times\wwB+\uuB\times\wwN+\jj\times\bbB+\jjB\times\bbN)',\\
\EEEB'&=(\uu\times\bbB+\uuB\times\bbN)'.
\end{align}
It is the one which comes closest to the quasi-kinematic
test-field method, because there
$\EEEB'=(\uu\times\bbB)'$.
Next, we substitute
$\meanBB$  by a test field, $\meanBBT$, and $\uuB$ and $\bbB$ by the test solutions, $\uuT$ and $\bbT$,
i.e.\
\begin{alignat}{2}
{\partial\uuT\over\partial t}
&=&\meanUU\times\wwT&+\uuT\times\meanWW
+\meanJJT\times\bb+\jj\times\meanBBT
\nonumber \\
&&&+\FFFT'+\nu\nabla^2\uuT-\nab\pT,
\label{duTdt_h}\\
{\partial\aaT\over\partial t}
&=&\meanUU\times\bbT&+\uu\times\meanBB^T
+\EEET'+\eta\nabla^2\aaT,
\label{daTdt_h}
\end{alignat}
where
\begin{align}
\FFFT'&=(\uu\times\wwT+\uuT\times\wwN+\jj\times\bbT+\jjT\times\bbN)',\\
\EEET'&=(\uu\times\bbT+\uuT\times\bbN)'.
\end{align}
For the mean electromotive force and force the ansatzes \eqref{Eansatz}
and \eqref{Fansatz} can be employed without change.
Note, however, that the tensors $\phiTens$ and $\psiTens$ now contain contributions from the Reynolds
stress created by $\uuB$, that is eventually, by $\meanBB$.
From the point of view of the tensorial structure of the relationship between $\meanBB$ and $\meanFFFF$ or $\meanEEEE$
the ansatzes \eqref{Eansatz} and \eqref{Fansatz} provide full generality as long as only the mean field and its first spatial derivative
are to be included. That's why there are no separate terms with the mean velocity or its gradient tensor. Instead 
the latter play the role of problem parameters and all transport coefficients can of course be determined as functions of them.
A separate task would consist in determining the tensors which appear in an analogous form of \eqref{E0ansatz} in place of the scalar coefficients
when a general anisotropic background is given. Then a test method  with respect to $\meanUU$ had to be tailored.

\section{Derivation of $\phiTens(k_z)$, $\psiTens(k_z)$ }
\label{phideriv}
Start with the stationary induction equation in  SOCA
\EQ
\eta\nabla^2\bbB+\curl(\uuN\times\meanBB)=\zervec.
\EN
Assume $\uuN = \urmsN\ff$ and $\bbN = \brmsN\ff$ with $\ff=\ff(x,y)$,
$\curl \ff = k_f \ff$, $\overline{\ff^2}=1$,
$\meanBB = \ithat{\BB} \ei{k_z z}$, 
and  $\ithat{B}_{x,y}=$ const, $\ithat{B}_z=0$.
Hence $\nabla^2 \ff = -k_f^2 \ff$.
Then we can make the ansatz $\bbB = \hat\bb(x,y) \ei{k_z z}$ with $\nabla^2 \hat\bb= -k_f^2 \hat\bb $ and get
\[
   {\bbB} = \rezip{\eta}\rezip{k_f^2+k_z^2} \big((\meanBB\cdot\nabla)  \uuN - {\rm i} k_z \,u_{0z}\meanBB \big)
\]
For the calculation of the mean force
\[
\FFFFB = \overline{\jjN\times\bbB} + \overline{\jjB\times\bbN}
\]
we need further 
\begin{align}
   \jjB &= \curl \bbB = \curl(\ei{k_z z} \hat{\bb}) = \ei{k_z z} ( \curl \hat{\bb} + {\rm i} k_z \zzz\times\hat{\bb} )  \\
          &= \frac{\kf}{\eta(\kf^2+k_z^2)}\big( (\meanBB\cdot\nabla) \uuN +  {\rm i} k_z (\meanBB\cdot\uuN) \zzz\big) + {\rm i} k_z \zzz\times\bbB .
\end{align}
Consequently
\begin{align*}
\FFFFB &=  k_f \overline{\bbN\times\bbB} + \overline{\jjB\times\bbN} \\
             &= \rezip{\eta} \rezip{k_f^2+k_z^2}\, \Big( \\
             & \phantom{=} {\rm i} k_z \overline{\big( \kf(\meanBB\cdot\uuN) \zzz + \kf u_{0z}\meanBB+ \zzz\times(\meanBB\cdot\nabla)\uuN\big) \times\bbN} \\
             & \phantom{=} +k_z^2 \,\overline{u_{0z}(\zzz\times\meanBB)\times\bbN}\,\Big)\\
             &= \rezip{\eta} \rezip{k_f^2+k_z^2}\, \Big( {\rm i} k_z \kf \overline{\big((\meanBB\cdot\uuN) \zzz + u_{0z}\meanBB\big)\times\bbN} \\
             & \phantom{=}  + {\rm i} k_z \overline{ b_{0z} (\meanBB\cdot\nabla)\uuN -  \zzz  \bbN\cdot(\meanBB\cdot\nabla)\uuN}\\
             & \phantom{=} +k_z^2(\overline{u_{0z}b_{0z}}\,\meanBB- \zzz \overline{u_{0z} \bbN}\cdot\meanBB)\,\Big)
\end{align*}
and with $\meanJJ={\rm i} k_z\zzz\times\meanBB$, that is, ${\rm i} k_z B_k = \epsilon_{ki3} \meanJ_i,\, k=1,2$,
\begin{align*}
\meanFFF_{\meanB i} &= \rezip{\eta}\rezip{k_f^2+k_z^2}\, \bigg( 
                                         \kf\big(\epsilon_{i3k}\epsilon_{ml3}\overline{u_{0m}b_{0k}} - \epsilon_{ijk}\epsilon_{kl3}\overline{u_{0z}b_{0j}}\big)\meanJ_l \\            
                &+ \epsilon_{lj3} \left( - \overline{b_{0z}\parder{u_{0i}}{x_j}} +  \delta_{i3}  \overline{\bbN\cdot\parder{\uuN}{x_j}}\, \right)\meanJ_l \\
           &\phantom{= }+k_z^2 \big(\, \overline{u_{0z}b_{0z}} \,\meanB_i - \delta_{i3} \overline{u_{0z} b_{0l}}\,\meanB_l\big)\bigg) .
\end{align*}
The tensors are hence
\begin{align*}
      \phi_{il} &=  \rezip{\eta}\frac{k_z^2}{k_f^2+k_z^2}\,\big (\,\overline{u_{0z}b_{0z} } \delta_{il} -  \overline{u_{0z}b_{0l} }  \delta_{i3}\big),\\
      \psi_{il} &=  \rezip{\eta}\rezip{k_f^2+k_z^2}\,\bigg( \kf( \overline{u_{0z}b_{0z}}\delta_{il}- \overline{u_{0z}b_{0l}}\delta_{i3})\\
                    &\phantom{=}  + \kf(1-\delta_{i3})\big(\overline{u_{0i}b_{0l}} - \delta_{il}(\overline{u_{01}b_{01}}+\overline{u_{02}b_{02}})\big)\\
                    &\phantom{=} + \epsilon_{lj3}\left (\, \overline{b_{0z} \parder{u_{0i}}{x_j}} -\, \overline{\bbN\cdot\parder{\uuN}{x_j}} \delta_{i3}\right) \bigg), \quad l\ne 3\\
      \phi_{i3} &= \psi_{i3} =0.
\end{align*}
For $k_z\rightarrow 0$ the tensor $\phiTens$ is proportional to $k_z^2$ thus the corresponding mean force expressed in physical space by a convolution $\itbreve{\phiTens}\circ\meanBB$, with $\itbreve{\phiTens}$ being the Fourier-backtransformed $\phiTens$, can be approximated by a term
$\propto\parderil{^2\meanBB}{z^2}$. For $k_z\gg\kf$, however,
the mean force can be represented by a term $\propto \meanBB$.
  
With Roberts geometry we have
for $\sigma=1$
\begin{align*}
  \phi_{11} &= \phi_{22} = 	\rezip{2\eta}\frac{k_z^2}{k_z^2+k_f^2} u_{0\rms}b_{0\rms}\,,\quad\psiTens=\zervec. 
\end{align*}
All other $\phi$ components vanish, too.

If, however, for the Roberts geometry $0\le\sigma<1$ the field $\ff$ has indeed yet the property $\nabla^2 \ff = -k_f^2 \ff$, but is no longer of Beltrami type.
Instead, we have
\[
  \curl \ff = \sigma \kf ( \ff + \frac{1-\sigma^2}{\sigma^2} f_z \zzz).
\] 
The tensor $\psiTens$ does not vanish any longer, but is now given by
\begin{align*}
  \psi_{11} &= -\rezip{\eta(k_z^2+k_f^2)} \frac{k_y^2(1-\sigma^2)}{\kf(1+\sigma^2)} u_{0\rms} b_{0\rms},\\
  \psi_{22} &= -\rezip{\eta(k_z^2+k_f^2)} \frac{k_x^2(1-\sigma^2)}{\kf(1+\sigma^2)} u_{0\rms} b_{0\rms},\\
  \psi_{12} &= \psi_{21} = 0.
\end{align*}

\section{Illustration of extracting a linear evolution equation from a
nonlinear one}
\label{Linear}

To illustrate the procedure of extracting a linear evolution equation
from a nonlinear problem, let us consider a simple quadratic ordinary
differential equation, $y'=y^2$, where a prime denotes
here differentiation.
We split $y$ into two parts, $y=\yN+\yL$, so we have
\begin{equation}
y^2=\yN^2+2\yN\yL+\yL^2. \label{nonlinGl}
\end{equation}
In the last two terms we can replace $\yN+\yL$ by $y$, 
so we have $2\yN\yL+\yL^2=(\yN+y)\yL$, which is now formally linear in $\yL$.
Here, $y$ corresponds to the solution of the `main run'.
Thus, at the expense of having to solve an additional nonlinear auxiliary
equation, $\yN'=\yN^2$, we have extracted a linear evolution equation
for $\yL$.
Altogether we have
\begin{equation}
\left\{
\begin{array}{rcl}
y'&=&y^2,\\
\yN'&=&\yN^2,\\
\yL'&=&(\yN+y)\yL,
\end{array}
\right.\label{linSys}
\end{equation}
where the last equation is linear in $\yL$.
Note, that the system \eqref{linSys} is 
exactly equivalent to \eqref{nonlinGl},
i.e.\ no approximation has been made.

\section{Derivation of \eqref{alpRoberts1} }
\label{PurelyMagneticRoberts}

Consider the stationary version of \eqref{du1dt} with $\FFFB'$ dropped (i.e. SOCA) and a uniform $\meanBB$, i.e., $\meanJJ=\zervec$
\EQ
\nu\nabla^2\uuB+\jj\times\meanBB=\zervec.
\label{app1}
\EN
Assume $\bb=\curl\aaaa,\,\dive\aaaa=0$, hence $\jj=-\nabla^2\aaaa$. We get
\EQ
\uuB =\aaaa\times\meanBB/\nu
\label{app2}
\EN
and further
\[
    (\overline{\uuB\times\bb})_i = \frac{1}{\nu}\epsilon_{ilm}\epsilon_{lkj}\overline{a_kb_m}\, \meanB_j = \alpha_{ij} \meanB_j
\]
that is,
\[\alpha_{ij} = (  \overline{\aaaa\cdot\bb}\, \delta_{ij} - \overline{ a_i b_j })/\nu.\]
Isotropy results in
\[ \alpha = \alpha_{ii}/3 = 2 \,\overline{\aaaa\cdot\bb }/{3\nu}. \]
For $\bb$ with Roberts geometry, however, we have $\alpha=\alpha_{11}=\alpha_{22} \ne \alpha_{33}$, hence
\[\alpha =  (  \overline{\aaaa\cdot\bb} + \overline{a_3b_3} )/2\nu = k_f (  \overline{\aaaa^2} + \overline{a_3^2 })/2\nu = 3\brms^2/{4 k_f \nu} \]
and with $\Lu=\brms/\eta k_f$
\EQ 
\alpha = \frac{3}{4}\, \brms\Lu / \Pm . \label{app3}
\EN

Adopt now a $\meanBB$ depending on $z$ only with
$\meanBB\propto {\rm e}^{\ii k_z z}$, but $\bb$, $\aaaa$ still independent of $z$.
Roberts geometry implies $\nabla^2 \aaaa = -k_f^2 \aaaa$ and $\nabla^2 \uuB = -(k_f^2+k_z^2)\uuB$. Inserting in \eqref{app1}
(with the term proportional to $\meanJJ$ omitted) yields
\[
   (k_f^2+k_z^2)\uuB = k_f^2 \aaaa\times\meanBB/\nu + \ldots
\]
and comparison with \eqref{app2} reveals that \eqref{app3} has only to be modified by the factor $ 1/\big(1+(k_z/k_f)^2\big)$.

\section{Derivation of $\alpha_\smk$ in fourth order approximation}
\label{alpmkDer}
We employ the iterative procedure
described, e.g., in R\"adler \& Rheinhardt (2007)
to obtain those contributions to $\EEEEB$ which are quadratic in $u_{0\rms}$ and $b_{0\rms}$ and 
expand for that purpose $\bbB$ and $\uuB$ into the series
\begin{align*}
\bbB &= \bbB^{(1)} + \bbB^{(2)} + \bbB^{(3)} + \ldots\,,\\
\uuB &= \uuB^{(1)} + \uuB^{(2)} + \uuB^{(3)} + \ldots
\end{align*}
with 
\begin{alignat*}{2}
&\eta\nabla^2\bbB^{(1)} &&= - \curl(\uuN\times\meanBB)\\
&\nu\nabla^2\uuB^{(1)} &&= - (\jjN\times\meanBB + \meanJJ\times\bbN) \\
&\eta\nabla^2\bbB^{(i)} &&= - \curl( \uuN\times\bbB^{(i-1)}  + \uuB^{(i-1)}\times\bbN )\\
&\nu\nabla^2\uuB^{(i)} &&= - ( \jjN\times\bbB^{(i-1)}  + \jjB^{(i-1)}\times\bbN )\,,\quad i=2,\ldots\\
\end{alignat*}
and
\[
 \EEEEB = \sum_{i=1}^\infty \left(\, \overline{ \uuN\times\bbB^{(i)} } + \overline{ \uuB^{(i)}\times\bbN } \,\right) = \sum_{i=1}^\infty \, \EEEEB^{(i)}
\]
In the following we assume $\meanBB$ to be uniform and $\uuN$, $\bbN$ to have Roberts geometry \eqref{robForce}.
The SOCA solutions $\bbB^{(1)}$ and $\uuB^{(1)}$ read
\[
  \bbB^{(1)} =  \rezip{\eta\kf^2} (\meanBB\cdot\nabla)\uuN,\quad \uuB^{(1)}  = \rezip{\nu\kf} \bbN\times\meanBB\,.
\]
From here on we switch to dimensionless quantities and set $\eta=\nu=1$, $k_x=k_y=1$, $\kf=\sqrt{2}$, $|\meanBB|=1$.
So we have
\begin{align*}
\bbB^{(1)} &= \frac{u_{0\rms}}{2}[\sin x \sin y, \cos x \cos y, -\sqrt{2} \sin x \cos y ] \\
 \uuB^{(1)} &=  \frac{b_{0\rms}}{2}[0,2\cos x \cos y, -\sqrt{2} \sin x \cos y]\\
\uuB^{(2)} &= \zervec\\
\bbB^{(2)} &= \rezip{8}\Big(-u_{0\rms}^2[\cos 2y, 0, \sqrt{2}\sin 2y] + \frac{b_{0\rms}^2}{2} \\
&[\cos 2y(\cos 2x+2),\sin 2y\sin 2x, \sqrt{2}\sin 2y(\cos 2x+3)] \Big).
\end{align*}
For $\bbB^{(3)}$ and $\uuB^{(3)}$ we present here only  those parts which eventually contribute to $\alpha_\smk$:
\begin{align*}
 \bbB^{(3)} &= \frac{u_{0\rms} b_{0\rms}^2}{32}[ \sin x \sin y, \cos x \cos y, -4\sqrt{2} \sin x \cos y ] \\
                    &\phantom{=} + \ldots\\
 \uuB^{(3)} &= \frac{u_{0\rms}^2 b_{0\rms}}{16}[ 0, \cos x \cos y, -\frac{\sqrt{2}}{2} \sin x \cos y  ] + \ldots \; .
\end{align*}
Finally,
\begin{align*}
 \EEEEB^{(3)}=\overline{\uuN\times\bbB^{(3)}} + \overline{ \uuB^{(3)}\times\bbN} 
                         = -u_{0\rms}^2 b_{0\rms}^2 \frac{\sqrt{2}}{64} + \ldots\,,
\end{align*}
i.e.\
\begin{align*}
 \alpha_\smk \approx  -u_{0\rms}^2 b_{0\rms}^2 \frac{\sqrt{2}}{64}  \,.          
\end{align*}
Note, that the contributions omitted in $ \EEEEB^{(3)}$ provide fourth order corrections to $\alpha_\sk$ and $\alpha_\sm$.
They result in dependences on $\Rm$ and $\Lu$ that are weaker than the
parabolic SOCA ones; see \figref{pRobMK_ubrmsDep_alp}.

\end{document}